\documentclass[twocolumn]{aastex701}
\usepackage{amssymb,amsmath,amsfonts,graphicx,epsf}
\usepackage{aas_macros}
\usepackage{lipsum}
\usepackage[dvipsnames]{xcolor}
\usepackage{bm}
\usepackage{subcaption}

\makeatletter
\newcommand{\oset}[3][0ex]{%
  \mathrel{\mathop{#3}\limits^{
    \vbox to#1{\kern-2\ex@
    \hbox{$\scriptstyle#2$}\vss}}}}
\makeatother
\newcommand{\tensorSymbol}[1]{{\oset[-0.2ex]{\leftrightarrow}{#1}}}
\graphicspath{{./}{figures/}}

\usepackage{hyperref}%
\hypersetup{
  colorlinks=true,       %
  linkcolor=MidnightBlue,%
  citecolor=MidnightBlue,%
  urlcolor=MidnightBlue  %
}

\begin{document}

\title{ Eccentric Binaries Accreting from Thin Disks: Orbital Evolution}

\correspondingauthor{Alexander J. Dittmann}
\email{dittmann@ias.edu}

\author[0000-0001-6157-6722]{Alexander J. Dittmann}
\affiliation{Institute for Advanced Study, 1 Einstein Drive, Princeton, NJ 08540, USA}
\altaffiliation{NASA Einstein Fellow}
\email{dittmann@ias.edu}

\author[0000-0001-9068-7157]{Geoffrey Ryan}
\affiliation{Perimeter Institute for Theoretical Physics, 31 Caroline St. N., Waterloo, ON, N2L 2Y5, Canada}
\email{gryan@perimeterinstitute.ca}

\author[0000-0002-5427-1207]{Luciano Combi}
\affiliation{Perimeter Institute for Theoretical Physics, 31 Caroline St. N., Waterloo, ON, N2L 2Y5, Canada}
\affiliation{Department of Physics, University of Guelph, Guelph, Ontario N1G 2W1, Canada}
\email{lcombi@perimeterinstitute.ca}

\begin{abstract} 
Circumbinary disks crucially affect the orbital and electromagnetic properties of binary systems across the universe, from stars in our galactic neighborhood to supermassive black hole binaries formed as the result of tumultuous galactic mergers. Previous simulations have focused nearly exclusively on thick accretion disks, appropriate for studying stellar binaries, and have found encouraging agreement with observations thereof. We present herein the first systematic study of eccentric binary systems accreting from thin disks, focusing on binary orbital evolution. Our main results are that (1) thinner disk not only drive binaries to rapidly inspiral, but also excite binary eccentricities at much higher rates; (2) while thick disks may drive binaries to a stable fixed point of $e\approx0.425$, thinner disks pump binary eccentricities to $e\gtrsim0.6$; (3) the range of near-zero eccentricities that are damped towards zero depends on both disk thickness and viscosity, thinner disks and those with $\alpha$ viscosities driving binaries towards circularity over a much narrower range of eccentricities. 
These differences follow largely from the effects of pressure support on accretion streams and shocks within the inner regions of the accretion flow. Our results suggest that accreting binary black holes should have high eccentricities well into the frequency range probed by pulsar timing arrays and space-based gravitational wave interferometers, affecting the spectrum and isotropy of the gravitational wave background. Our results also suggest that circumbinary disks may play an important role in shaping the orbits of close binary stars, but much less so those of wider binaries. 

\end{abstract}


\section{Introduction}
Binary systems, from stellar pairs within our own Milky Way to binary supermassive black holes in the distant reaches of the Universe, are sculpted by interaction with their gaseous environments. Circumbinary disks may shape nascent stellar binaries, shortly after their collapse from molecular clouds \citep[e.g.,][]{kant1755allgemeine,2023ASPC..534..275O}, or later in life after asymptotic-giant branch (AGB) stars shed their outer layers \citep[e.g.,][]{2006A&A...448..641D,2009A&A...505.1221V}.\footnote{Compared to accreting black holes, stellar binaries have the aesthetic advantage of having resolvable circumbinary disks, such as that of GG Tauri A, L1448 IRS3B, and many others \citep[e.g.,][]{2016Natur.538..483T,2019Sci...366...90A,2020A&A...639A..62K}.} Supermassive black hole binaries, formed in the aftermath of galactic mergers \citep[e.g.,][]{1980Natur.287..307B,2013CQGra..30x4008M}, are thought to accrete from the gas abounding in post-merger galactic centers \citep[e.g.,][]{1991ApJ...370L..65B,2005ApJ...620L..79S}, which can accelerate their inspirals and may reveal their spatially unresolvable binarity \citep[e.g.,][]{1980Natur.287..307B,2023ARA&A..61..517L,2023arXiv231016896D}.

Most binaries will form with nonzero eccentricity unless circularized \citep[for example, by tides, e.g.,][]{1995A&A...296..709V,2018ApJ...867....5P}. Moreover, the eccentricities of binary black holes are typically excited further by gravitational interactions with wandering stars \citep[e.g.,][]{2008ApJ...686..432S,2024MNRAS.532..295F}. Understanding the influence of circumbinary disks on eccentric binaries is thus essential to interpreting observations of stellar binaries \citep[e.g.,][]{2007A&A...463..683S,2019MNRAS.489.5822E,2022ApJ...933L..32H}. Such knowledge will also be crucial for modeling supermassive black hole binary populations as probed by pulsar timing arrays \citep[e.g.,][]{2016ApJ...817...70T,2017MNRAS.470.1738C,2021MNRAS.501.2531S} and interpreting the electromagnetic signatures of putative binaries in electromagnetic surveys \citep[e.g.,][]{2022PhRvD.106j3010W,2024ApJ...977..244D}.

Previous simulations of eccentric accreting binaries have found that nearly circular binaries are driven towards circularity, that moderately eccentric binaries are driven to higher eccentricities, and that highly eccentric binaries are circularized \citep[e.g.,][]{2019ApJ...871...84M,2021ApJ...909L..13Z,2021ApJ...914L..21D}; such simulations predict stable binary eccentricities of $e\sim0$ and $e\sim0.4$,\footnote{ \citet{2023MNRAS.522.2707S} found the upper stable eccentricity to decrease with binary mass ratio ($q\equiv m_2/m_1\leq1$) from $e\sim0.45$ at $q=1$ to $e\sim0.3$ at $q=0.1$, and that circular orbits were unstable.} in qualitative agreement with observations of post-AGB binaries \citep[e.g.,][]{1998A&A...332..877J,2018A&A...620A..85O}. However, some Lagrangian simulations have hinted at both higher and lower stable binary eccentricities depending on the gas cooling rate, (self-gravitating) disk mass, and binary mass ratio \citep[e.g.,][]{2011MNRAS.415.3033R,2024MNRAS.534.3448B,2024A&A...688A.174F}.  

Previous systematic studies have focused almost exclusively on thick accretion disks, with aspect ratios $H/R\sim0.1$ (or characteristic azimuthal Mach number $\mathcal{M}=10$), appropriate to some stellar binaries \citep{2021ApJ...909L..13Z,2021ApJ...914L..21D,2023MNRAS.522.2707S}.\footnote{A notable exception is the study of \citet{2025MNRAS.537.2422P} on circumbinary protoplanetary disks, which simulated low-viscosity disks with a variety of aspect ratios around binaries with eccentricities $0\leq e\leq0.4$ and mass ratio $q=0.26$. Unfortunately, those simulations did not resolve accretion and their results suffered accordingly from large uncertainties. Nevertheless, their results exhibit many common trends with those of this study.} However, the disks of gas accreting onto supermassive black holes are thought to be quite thin, with $H/R\sim10^{-2}$ \citep[e.g.][]{2002apa..book.....F}. Furthermore, even mildly thin disks with $\mathcal{M}\sim20-30$ cause binaries to evolve in strikingly different ways than thicker $\mathcal{M}\sim10$ disks, typically driving rapid inspirals rather than slow outspirals \citep{2020ApJ...900...43T,2022MNRAS.513.6158D,2024ApJ...967...12D,2025ApJ...984..144T}. Moreover, the aforementioned thin-disk studies were limited to circular binaries, and the influence of disk thickness on eccentric binaries has been largely unexplored.  

Motivated thus, we have carried out a comprehensive suite of simulations, focusing on equal-mass eccentric binary orbital evolution, varying primarily the binary eccentricity and disk thickness, but also varying the magnitude and functional form of the viscosity, and investigating the importance of reaching inflow equilibrium in a subset of our simulations. Section \ref{sec:orbits} reviews the basics of binary orbits and the evolution thereof, describes the equations of vertically integrated hydrodynamics applied in this study, and describes our diagnostics and initial conditions. In particular, Section \ref{sec:orbitalDiagnostics} derives a number of expressions that clarify the causes of binary orbital evolution for arbitrary eccentricities and mass ratios. We present the results of our simulations in Section \ref{sec:results} and explore therein the physical mechanisms by which thinner disks drive binaries to higher eccentricities. In Section \ref{sec:discussion} we discuss the implications of our findings on the gravitational waves emitted from supermassive black hole binaries and discuss the limitations of our work. We summarize our results in Section \ref{sec:summary} and tabulate the results of our simulations in Appendix \ref{app:theTable}.

\section{Preliminaries}\label{sec:orbits}
\subsection{Orbits}
Given a binary with total mass $m$, separation vector $\mathbf{r}\equiv\mathbf{r}_2-\mathbf{r}_1$, and relative velocity vector $\mathbf{v}=\mathbf{v}_2-\mathbf{v}_1$, the specific energy ($\mathcal{E}$) and specific angular momentum vector ($\mathbf{h}$) of the binary are
\begin{equation}
\mathcal{E}=\frac{1}{2}\mathbf{v}\cdot\mathbf{v} - \frac{Gm}{|\mathbf{r}|},~~~\mathbf{h}=\mathbf{r}\times\mathbf{v},
\end{equation}
which in turn define the binary semi-major axis ($a$) and scalar eccentricity ($e$)
\begin{equation}\label{eq:orbdef}
a=-\frac{Gm}{2\mathcal{E}},~~~e^2=1-\frac{h^2}{Gma}=1+\frac{2h^2\mathcal{E}}{G^2m^2},
\end{equation}
where $h\equiv\sqrt{\mathbf{h}\cdot\mathbf{h}}$ and $G$ is the gravitational constant. From the definition of the semi-major axis, we can calculate the orbital frequency, or mean motion, $n=G^{1/2}m^{1/2}a^{-3/2}$, for the binary or for a given fluid element. 
Differentiating Equation \ref{eq:orbdef} in time enables us to relate changes of the energy and angular momentum of a binary to changes in its orbital elements:
\begin{align}
\frac{\dot{a}}{a} &= \frac{\dot{m}}{m} - \frac{\dot{\mathcal{E}}}{\mathcal{E}},\\
\frac{e\dot{e}}{1-e^2} &= \frac{\dot{m}}{m} - \frac{1}{2}\frac{\dot{\mathcal{E}}}{\mathcal{E}} - \frac{\dot{h}}{h}.
\end{align}
The specific energy and angular momentum of the binary are altered by external forces and the corresponding accelerations, $\mathbf{a}_e$, as well as changes in the binary mass due to accretion:
\begin{align}
\dot{\mathcal{E}} = \mathbf{a}_e\cdot\mathbf{v} - \frac{G\dot{m}}{|\mathbf{r}|}\\
\dot{h} = \mathbf{r}\times\mathbf{a}_e.
\end{align}

In the following we will consider binaries with their orbital planes aligned with the midplane of a surrounding thin accretion disk; we will approximate the evolution of the accretion disk by solving the vertically integrated equations of hydrodynamics and following only the in-plane dynamics. 

\subsection{Hydrodynamics}
The equations of 2D vertically integrated hydrodynamics are given by  
\begin{align}
\partial_t\Sigma + \bm{\nabla}\!\cdot\!(\Sigma\mathbf{V}) &= S_\Sigma, \label{eq:continuity} \\
\partial_t(\Sigma\mathbf{V})\! +\! \bm{\nabla}\!\cdot\!(\Sigma\mathbf{V}\mathbf{V}+\Pi\tensorSymbol{I}\!- 2 \Sigma \nu \tensorSymbol{\sigma})
&= -\Sigma\bm{\nabla}\Phi + \mathbf{S}_{p},
\label{eq:momentum}
\end{align}
where $\Sigma$ is the fluid surface density, $\Pi$ is the vertically-integrated pressure, $\mathbf{V}$ is the mass-weighted vertically-integrated fluid velocity, $\nu$ is a (possibly time- and space-dependent) kinematic viscosity, and $\tensorSymbol{I}$ and $\tensorSymbol{\sigma}$ are the identity and velocity shear tensors, respectively. The gravitational potential of the binary is given by $\Phi$, and the density and momentum sink terms stemming from accretion onto the binary are given by $S_\Sigma$ and $\mathbf{S}_p$ respectively. We employ a locally isothermal equation of state given by 
$\Pi=c_s^2\Sigma$, where the sound speed is specified by $c_s^2 = -\Phi/\mathcal{M}^{-2}$ where $\mathcal{M}$ is a characteristic azimuthal Mach number characterizing the thickness and degree of pressure support within the disk. 

The gravitational potential of the binary is $\Phi = \sum_i \Phi_i$, where we artificially soften the potential of each particle on a scale $\epsilon_s$ according to 
\begin{equation}
\Phi_i = \frac{-Gm_i}{\sqrt{|\mathbf{x}-\mathbf{r}_i|+\epsilon_s^2}}.
\end{equation}
The surface density sink term is given by 
\begin{equation}
S_\Sigma = -\gamma\Omega_0\Sigma\sum_i s_i(|\mathbf{x}-\mathbf{r}_i|),
\end{equation}
where $\gamma$ is a scalar that sets the mass removal rate relative to the binary orbital frequency $\Omega_0$ and $s_i$ is a function that limits accretion to the immediate vicinity of each particle. We set $s_i=\exp{(-[|\mathbf{x}-\mathbf{r}_i|/r_s]^4)}$, where $r_s$ is the characteristic sink radius. The momentum sink term is given by
\begin{equation}
\mathbf{S}_p = -\gamma\Omega_0\Sigma\sum_i s_i \mathbf{V}^*_i.
\end{equation}
We chose $\mathbf{V}_i^*=(\mathbf{V}-\mathbf{v}_i)\cdot\hat{\mathbf{r}}_i\hat{\mathbf{r}}_i+\mathbf{v}_i$, where $\hat{\mathbf{r}}_i$ is a radial basis vector in a polar coordinate system centered on the accreting particle; this ``torque-free'' treatment eliminates sink-dependent torques on the gas, correctly reproduces steady disk profiles, and makes simulation results much less dependent of the rate of mass removal $\gamma$ \citep[e.g.,][]{2020ApJ...892L..29D,2021ApJ...921...71D,2024ApJ...967...12D}.

\subsection{Disk-Driven Orbital Evolution}\label{sec:orbitalDiagnostics}
We can quantify the interaction between the binary and its accretion disk through the aforementioned sink terms, defining the accretion rate onto each member of binary as 
\begin{equation}
\dot{m}_i = -\int dA\, S_{\Sigma,i},
\end{equation}
the force on each member of the binary due to accretion as 
\begin{equation}
\mathbf{F}_{a,i} = -\int dA\, \mathbf{S}_{p,i},
\end{equation}
and the gravitational force on the binary as 
\begin{equation}
\mathbf{F}_{g,i} = \int dA\, \Sigma \nabla \Phi_i.
\end{equation}
The external accelerations acting on the binary are thus $\mathbf{a}_{a}=(\mathbf{F}_{a,2}-\mathbf{v}_2\dot{m}_2)/m_2-(\mathbf{F}_{a,1}-\mathbf{v}_1\dot{m}_1)/m_1$ (due to accretion) and 
$\mathbf{a}_{g}=\mathbf{F}_{g,2}/m_2-\mathbf{F}_{g,1}/m_1$ (due to gravity).
The form we have chosen for $\mathbf{V}^*_i$ ensures that accretion contributes only to the \textit{orbital} angular momentum of each sink particle rather than its spin; this is appropriate for accretion on scales well below the resolution limit of the grid, but not those with surfaces or horizons at similar scales to the sink radius. See \citet{2021ApJ...921...71D} for additional details.

If accretion is well-resolved, then the gas around each sink particle moves with the same average specific specific momentum as the object onto which it accretes and $\langle\mathbf{a}_a\rangle=0$, such that $\mathbf{a}_e\approx\mathbf{a}_g$; this holds very well when accretion disks form around each member of the binary but breaks down when the binary experiences a headwind \citep[e.g.,][]{2024ApJ...964...61D,Dittmann:2025aup}. In this case, the definitions of $\mathcal{E}$ and $\mathbf{h}$ yield that the specific angular momentum of the binary is only affected by gravitational interactions while its specific energy is affected by both gravitational interactions and accretion:
\begin{align}
\dot{\mathcal{E}}&=\mathbf{v}\cdot\mathbf{a}_g-\frac{G}{|\mathbf{r}|}\dot{m},\\
\dot{\mathbf{h}}&=\mathbf{r}\times\mathbf{a}_g.
\end{align}
Because we consider in-plane binaries in this paper, we can write  $\dot{h}=|\dot{\mathbf{h}}|$ and the evolution of the binary orbital elements is thus given by
\begin{align}
\frac{\dot{a}}{a}&=-\frac{\mathbf{v}\cdot\mathbf{a}_g}{\mathcal{E}} + \frac{\dot{m}}{m}\left(1-\frac{2a}{|\mathbf{r}|}\right)\\ \label{eq:dedm}
\frac{e\dot{e}}{1-e^2}&=-\frac{|\mathbf{r}\times\mathbf{a}_g|}{h} -\frac{\mathbf{v}\cdot\mathbf{a}_g}{2\mathcal{E}}+\frac{\dot{m}}{m}\left(1 - \frac{a}{|\mathbf{r}|}\right).
\end{align}
It is worth dwelling upon the preceding expressions for a moment. Note that their derivation at no point required assuming a particular binary mass ratio or worrying about the amount of accretion onto one object or the other, and they are thus valid for all mass ratios and have no dependence on whether one object or the other accretes preferentially. It is also worth noting that eccentric binaries, unlike circular ones, are affected non-trivially by accretion: the changes in their semi-major axes and eccentricities depend on where over the course of an orbital period the binary accretes most. Additionally, both the left- and right-hand sides of Equation (\ref{eq:dedm}) are zero when $e\rightarrow0$.

\subsection{Numerical Methods}\label{sec:numerics}
We used the moving-mesh finite volume code \texttt{Disco} \citep{2016ApJS..226....2D}\footnote{Specifically, we used the version \url{https://github.com/ajdittmann/Disco}, which includes additional optimizations and in-situ diagnostics. Unlike the original version of \texttt{Disco} described in \citet{2016ApJS..226....2D} which only treats viscosity correctly when $\nu\Sigma={\rm const.}$, we included the full set of viscous terms in the momentum equation as described in \citep{2021ApJ...921...71D}.} to solve the equations of vertically integrated hydrodynamics (Equations \ref{eq:continuity}--\ref{eq:momentum}). We solved the aforementioned equations in a cylindrical polar $(R,\phi)$ coordinate system centered on the binary center of mass. The radial grid spacing was linear for $0\leq R<a$ and logarithmic for $a\leq R \leq 30a$. The domain was decomposed into $N_R=1152$ annuli, and the number of cells in each annulus chosen to maintain $r\Delta \phi \approx \Delta R$ as closely as possible. The cell size within $R<a$ was $\Delta R = 0.00382a$, resolving the binary semi-major axis by 261 cells. The highest binary eccentricity we considered was $e=0.6$, in which case the binary pericenter distance was resolved by about 104 cells. 

We set $r_s=\epsilon_s=0.025$, so both were sufficiently resolved by our grid spacing yet small enough compared to the binary semi-major axis (and separation at pericenter) to have little effect on our results \citep[e.g.,][]{2024ApJ...970..156D}. Accordingly, we set $\gamma=7.5$ so that the sink timescale was moderately shorter than the viscous timescale at $r_s$. Our simulations employed the Harten-Lax-van Leer-Contact approximate Riemann solver \citep{1994ShWav...4...25T} and a second-order total-variation-diminishing spatial scheme \citep{1979JCoPh..32..101V,2000JCoPh.160..241K}. Each annulus of the computational domain moved with the average azimuthal velocity of the fluid within that annulus to reduce numerical dissipation. Our simulations used second-order Runge-Kutta time stepping \citep{1998MaCom..67...73G}. We typically set the Courant–Friedrichs–Lewy timestep factor to $\sim0.5$, but lowered it to $0.1$ for simulations of the highest-eccentricity binaries accreting from thin ($\mathcal{M}\sim30$) disks to stably resolve gas flows around the binary near pericenter.

\subsubsection{Initial Conditions}

The initial surface density profile in our simulations was given by
\begin{equation}\label{eq:initSig}
\Sigma=\varsigma + (1-\varsigma)e^{-\left(\frac{R}{r_c}\right)^{-\xi}}\left[\!1\!-\!\ell_0\left(\frac{R}{a}\right)^{\!\!-1/2}\right]\!\!\left(\frac{R}{a}\right)^{\!\!-p},
\end{equation}
where $r_c=2.5a$ sets the initial radius of a cavity around the binary, $\xi=12$ sets the steepness of this cavity, and $\varsigma=10^{-5}$ is a constant which sets the minimum density for our initial condition. The exponent
$p$ controls the surface density profile far from the binary and is set to $p=0$ for our simulations employing globally constant kinematic viscosities, but $p=1/2$ for a limited set of simulations using spatially-dependent ``$\alpha$'' viscosities. The constant $\ell_0$ sets the initial angular momentum flux through the disk at large distances from the binary: when $\ell_0$ differs from the ratio the rates of angular momentum and mass transfer to the binary ($\dot{J}/\dot{m}$), accretion can be temporarily enhanced or suppressed onto the binary until the system approaches equilibrium. We typically set $\ell_0=0$, but in some cases carefully chose $\ell_0$ close to the steady-state value (inferred from an initial $\ell=0$ simulation) to assess the impact of our initial simulations not reaching equilibrium. 
 
The initial radial fluid velocity in our simulations was set to
\begin{equation}\label{eq:broken}
v_r = -\frac{3\nu}{2\mathcal{R}},
\end{equation}
where $\mathcal{R}=\sqrt{R^2+\epsilon_s^2}$ was chosen to avoid divergence at the origin. 
The initial azimuthal velocity ($v_\phi=R\Omega$) was specified through approximate hydrostatic equilibrium, taking into account both pressure gradients and the zero-eccentricity binary quadrupole moment, 
\begin{equation}
\Omega^2=\Omega_K^2(\mathcal{R})\left(1+\frac{3a_b^2}{4\mathcal{R}^2}\right)+\frac{1}{\mathcal{R}\Sigma}\frac{d\Pi}{dr},
\end{equation}
where $\Omega_K^2(R)=GmR^{-3}$. 

\begin{figure*}
\includegraphics[width=\linewidth]{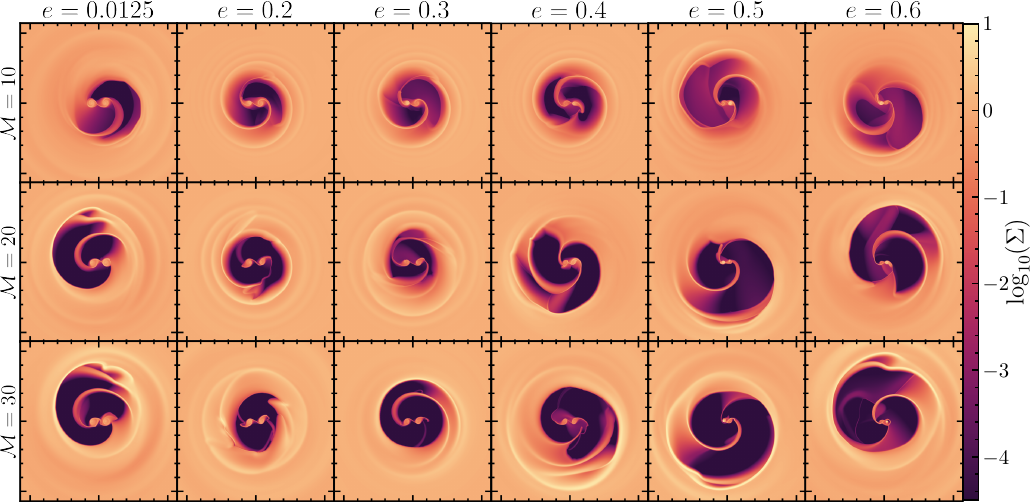}
\caption{Snapshots of the circumbinary disk surface density for a sample of the Mach numbers and binary eccentricities studied in this work. The binaries are shown here at pericenter. Tick markings are spaced in increments of the binary semi-major axis. Each case illustrates the fluid after 1800 orbital periods of the binary.}
\label{fig:surface}
\end{figure*}

\subsubsection{Diagnostics}
Although the \textit{specific} energy and angular momentum of the binary are useful by virtue of their direct connection to the binary semi-major axis and eccentricity, the conservation of angular momentum of the system makes it more natural to measure the rates of change of the binary angular momentum $J$ and energy $E$ from the simulation.\footnote{The conversion from specific to total quantities is, however, trivial given that $E=\mu\mathcal{E}$ and $J=\mu h$, where $\mu=m_1m_2/m$ and for the equal-mass binaries considered here $\dot{\mu}/\mu = \dot{m}/m$.} Thus, throughout each simulation, we record time averages of the torques on the binary due to accretion
\begin{equation}
\dot{\mathbf{J}}_{a} = -\int dA\, \mathbf{x}\times\mathbf{S}_{p},
\end{equation}
and due to gravitational interactions with the accretion disk
\begin{equation}\label{eq:gravtorque}
\dot{\mathbf{J}}_{g} = \int dA\, \mathbf{x}\times \left(\Sigma \nabla \Phi\right).
\end{equation}
Decomposing the binary energy $E$ into kinetic ($K$) and potential ($U$) components, we can compute the contributions to the kinetic energy of the binary energy similarly as
\begin{equation}
\dot{K}_{a} = -\int dA\, \mathbf{v}\cdot\mathbf{S}_{p},
\end{equation}
and due to gravitational interactions with the accretion disk
\begin{equation}\label{eq:gravpower}
\dot{K}_{g} = \int dA\, \mathbf{v}\cdot \left(\Sigma \nabla \Phi\right),
\end{equation}
and the change in the potential energy of the binary as
\begin{equation}
\dot{U} = -2G\mu\frac{\dot{m}}{|\mathbf{r}|}.
\end{equation}
We calculate time averages of each aforementioned quantity every time step, consistently with our Runge-Kutta time-stepping. We also calculate time-averages of the gravitational contributions to the energy and angular momentum of the binary as functions of radius, 
\begin{align}\label{eq:djdr}
\frac{d\dot{\mathbf{J}}_{g}}{dr} &= \int d\phi\, \mathbf{x}\times \left(\Sigma \nabla \Phi\right)\\ \label{eq:dkdr}
\frac{d\dot{K}_{g}}{dr} &= \int d\phi\, \mathbf{v}\cdot \left(\Sigma \nabla \Phi\right),
\end{align}
which can be used to understand which regions or aspects of the accretion disk drive particular trends in binary orbital evolution. For example, the above quantities can be combined linearly, following Equation (\ref{eq:dedm}),  to define
\begin{equation}\label{eq:dedr}
\frac{d\dot{e}_g}{dr}=\frac{1-e^2}{e}\left(-\frac{1}{J}\frac{d\dot{J}_{g}}{dr} -\frac{1}{2E}\frac{d\dot{K}_{g}}{dr}\right),
\end{equation}
which quantifies how various regions of the accretion disk affect the binary eccentricity.

\section{Results} \label{sec:results}
Our primary survey focused on disks with characteristic Mach numbers $\mathcal{M}\in\{10,20,30\}$ (or equivalently aspect ratios $H/r\in\{0.1,0.05,0.0\bar{33}\}$), over a range of eccentricities $e\in[0.0125,0.6]$, employed a globally constant kinematic viscosity $\nu=10^{-3}$, and set $\ell_0=0$; we discuss this survey in Sections \ref{sec:morphology} and \ref{sec:orbital} to provide a general overview of our results. Most importantly, the steady state binary eccentricity of $e\sim0.4$ found by many previous studies at $\mathcal{M}=10$ (which we reproduce) does not hold for thinner accretion disks, which instead drive binaries towards higher eccentricities $e>0.6$. This initial survey also suggested that while circular orbits are stable, with a basin of attraction up to $e\sim0.05$ at $\mathcal{M}=10$, near-circular binaries accreting from thinner disks are driven towards higher eccentricities much more easily, from eccentricities as low as $e=0.00625$ at $\mathcal{M}=30$. 

To investigate the causes and assess the generality of these results, we also conducted a series of targeted simulations, varying the kinematic viscosity (in magnitude and function form) in addition to sampling more densely in Mach number. We discuss high-eccentricity behavior in Section \ref{sec:attractors} and whether or not near-circular orbits are circularized in Section \ref{sec:nearcirc}. We provide a quantitative record of all of our orbital evolution results in Appendix \ref{app:theTable}.

\subsection{Disk Morphology}\label{sec:morphology}
Circumbinary disks differ enormously from accretion disks around single objects because of the extent to which the binary disrupts --- rather than perturbs --- the disk. The gravitational forcing of the binary on the disk is typically extremely nonlinear (in the sense that the scale hight of the disk is much smaller than the length scale associated with the tidal potential of the binary) and the disk experiences order-unity deviations from its time- and/or azimuthally-averaged state over a range of timescales. Binaries, even circular ones, typically excite noticeable eccentricities in their accretion disks and open deep cavities in their central regions \citep[e.g.,][]{2008ApJ...672...83M,2012ApJ...749..118S,2020MNRAS.499.3362R,2022MNRAS.513.6158D}. Moreover, the inner regions of circumbinary accretion disks $r\lesssim 4 a$ are highly variable and frequently re-arranged by strong shocks, even their azimuth- and orbit-averaged properties varying considerably orbit-to-orbit. 

\begin{figure}
\includegraphics[width=\linewidth]{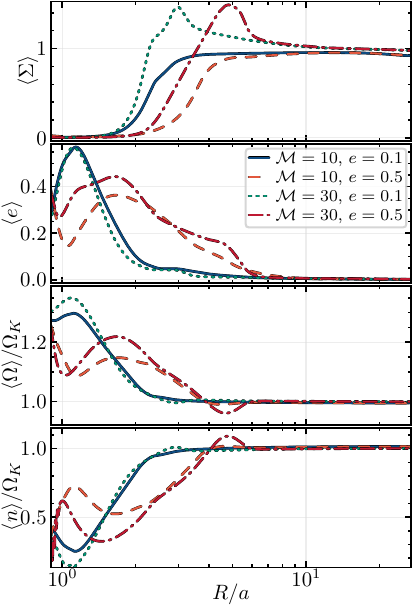}
\caption{Azimuthally integrated fluid profiles for Mach 10 and 30 disks around binaries with eccentricities of 0.1 and 0.5., time-averaged in each case over the final 1200 orbits of each simulation. The first and second rows plot the average disk surface density and eccentricity, demonstrating the mild eccentricities of the circumbinary disk and the even more pronounced eccentricity of fluid within the cavity. The bottom two rows plot the average angular and orbital frequencies of the fluid throughout the disk, normalized by the Keplerian frequency $\Omega_K\equiv (R/a)^{-3/2}$. All averages except that of the surface density were weighted by mass.}
\label{fig:diskProf}
\end{figure}

To illustrate the aforementioned characteristics, we present snapshots of the disk surface density for a wide variety of disk Mach numbers and binary eccentricities in Figure \ref{fig:surface}, and time- and azimuth-averaged fluid surface density, eccentricity, angular frequency, and mean motion in Figure \ref{fig:diskProf} for four representative disks. Salient trends visible in Figure \ref{fig:surface} include the overdense ring around the edge of the cavity and the underdensity of the cavity itself becoming more pronounced in thinner disks as streams of gas flung away from the binary shock against the cavity edge more strongly. The cavity itself tends to be the smallest around $e\sim0.2$ binaries, largest around $e\sim0.5$ binaries. 

Thinner disks have less support from pressure gradients against gravity, and lower sound speeds, leading to stronger shocks. 
These strong shocks excite larger eccentricities in the inner edge of the circumbinary disk \citep{2022MNRAS.513.6158D} and weakened pressure gradients leave streams thinner, as illustrated in Figure \ref{fig:surface}; these thinner streams are less able to accrete onto the binary, and in the well-studied case of circular binaries are known to help binaries rapidly inspiral \citep[see, for example,][]{2022MNRAS.513.6158D,2025ApJ...984..144T}. As we will illustrate below, the thinness of these streams also helps spur binaries towards higher eccentricities.

\begin{figure}
\includegraphics[width=\linewidth]{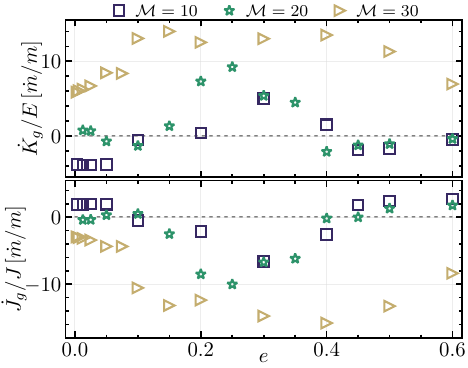}
\caption{Time-averaged measurements of the gravitational power and torque exerted on binaries by their disks, normalized by the energy and angular momentum of the binary respectively, in units of $\dot{m}/m$. We focus here on binaries with eccentricities $e\in[0.0125,0.6]$ and disks with Mach numbers $\mathcal{M}\in\{10,20,30\}$. All time-averaging was performed over the final 1200 binary orbits of each simulation.}
\label{fig:gravitationalScatter}
\end{figure}

\begin{figure}
\includegraphics[width=\linewidth]{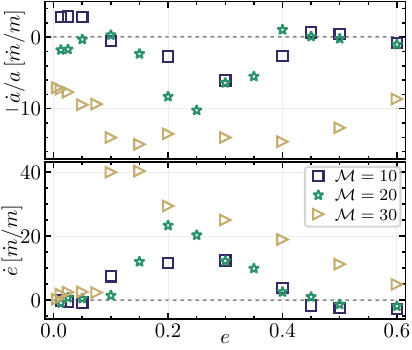}
\caption{Time-averaged measurements of the rates of change of binary semi-major axes and eccentricities for the same simulations presented in Figure \ref{fig:gravitationalScatter}. Notably, while thicker disks can drive binaries to eccentricity equilibria near $e\sim0.45$, thinner disks drive binaries towards substantially higher eccentricities, $e>0.6$. All time-averaging was performed over the final 1200 binary orbits of each simulation.}
\label{fig:orbitalScatter}
\end{figure}

\begin{figure*}
\includegraphics[width=\linewidth]{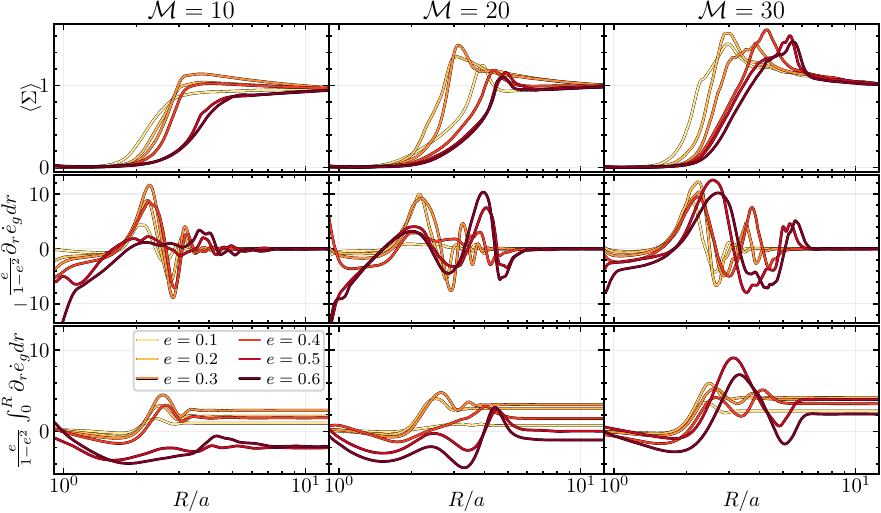}
\caption{Profiles of the disk surface density (top row), the contribution of the disk to the eccentricity evolution of the binary as a function of radius (middle row, see Equation (\ref{eq:dedr})), and its integral (bottom row), each integrated in azimuth and averaged over the final 20 orbits of each simulation. Note that we have rescaled $d\dot{e}_g/dr$ by a factor of $e/(1-e^2)$ to highlight the role of the disk rather than the changing properties of the binary. In most cases the inner regions of the accretion flow surrounding the binary itself (at $R<a$) contribute negligibly to the evolution of the binary. Typically, the streams of gas flowing towards and being jettisoned away from the binary and the inner edge of the circumbinary cavity govern the orbital evolution of the binary.}
\label{fig:eccDiskProf}
\end{figure*}

Visually, the cavity around each binary is clearly eccentric. Figure \ref{fig:diskProf} makes this more quantitative in a few cases, illustrating that the inner edge of the accretion disk can become moderately eccentric, to a degree which increases with Mach number; the fluid within the cavity tends to be considerably more eccentric, reaching mass-weighted average eccentricities in excess of 0.5. The angular velocity profile of the gas deviates from Keplerian only very near the binary and well within the cavity, generally as expected due to the higher-order multipole moments of the binary potential. 

Crucially, the typical semi-major axis of fluid elements within a given circumbinary disk cavity is nearly constant, and thus the typical orbital period (and mean motion) of the fluid throughout the cavity is approximately that of the cavity edge. Thus, despite the azimuthal velocity of fluid within the cavity exhibiting a nearly (or super) Keplerian profile with radius, the actual mean motion of fluid at a given radius within the cavity is usually much lower (and its orbital period much longer) than one would expect for a circular accretion disk. 
 
\subsection{Orbital Evolution}\label{sec:orbital}
We display the normalized power and torque delivered to the binaries in a variety of simulations by their gravitational interaction with the accretion disk in Figure \ref{fig:gravitationalScatter} (the accretion component has a much smaller effect, see Section \ref{sec:orbitalDiagnostics} and Appendix \ref{app:theTable}). While these two quantities largely track one another, especially at low binary eccentricities, it is subtle differences between $\dot{J}/J$ and $\dot{K}_g/2E$ that cause the binary eccentricity to evolve, as shown by Equations (\ref{eq:dedm}) and (\ref{eq:dedr}). The corresponding rates of change of the binary semi-major axis and eccentricity are shown in Figure \ref{fig:orbitalScatter}, where the evolution of the semi-major axis directly follows from the power delivered to the binary (and its change in mass).

For nearly-circular binaries ($e=0.0125$), we essentially reproduce the values for $\dot{a}/a$ measured in other works for circular binaries, confirming that they tend to expand when accreting from $\mathcal{M}\sim10$ disks but shrink when accreting from $\mathcal{M}\gtrsim15$ disks \citep[e.g.,][]{2020ApJ...900...43T,2022MNRAS.513.6158D}. However, deviations appear at even very slight eccentricities $e\sim0.025-0.05$. As noted previously \citep[e.g.,][]{2021ApJ...914L..21D,2023MNRAS.522.2707S}, at $\mathcal{M}=10$ binaries may either expand or contract depending eccentricity; but at $\mathcal{M}=20$ the windows of outspiral become more narrow, and by $\mathcal{M}=30$ binaries inspiral regardless of their eccentricity, and do so much more quickly than at lower Mach numbers. 

We reproduce the general result found at $\mathcal{M}=10$ that there is a fixed point binary eccentricity near $e\sim0.4$, $\dot{e}$ positive below and negative above \citep{2021ApJ...909L..13Z,2021ApJ...914L..21D,2023MNRAS.522.2707S}.\footnote{Let us not overstate the generality of this result: \citet{2021ApJ...914L..21D} found a fixed point around $e\approx0.39$, \citet{2021ApJ...909L..13Z} found a fixed point around $e\approx0.45$, and \citet{2023MNRAS.522.2707S} found that there should be a fixed point somewhere between $e=0.4$ and $e=0.5$.} The differences between $\mathcal{M}=10$ and $\mathcal{M}=20$ disks in this regard are very minor: the stable eccentricity value resides closer to $e\approx0.425$ at $\mathcal{M}=10$ and $e\approx0.0475$ at $\mathcal{M}=20$. 
This picture changes dramatically when binaries accrete from thinner disks: over the range of eccentricities surveyed here at $\mathcal{M}=30$, we find that eccentricity is always driven upwards, with extrapolation suggesting that a fixed point may reside near $e\approx0.7$. Thinner disks appear to more readily excite eccentricities quite generally, in addition to driving much faster orbital shrinkage. 

Despite the magnitude of $\dot{e}$ changing dramatically with disk thickness, some trends hold somewhat generally: $\dot{e}$ tends towards zero at large eccentricities, and the general shape of $\dot{e}(e)$ is sculpted by the overall prefactor $(1-e^2)/e$ regardless of the torque and power exchanged between disk and binary. But overtop this necessary trend, accretion flows with different Mach numbers drive remarkably different orbital evolution. To investigate the origins of these differences with Mach number, we plot in Figure \ref{fig:eccDiskProf} time averages of the surface density and $\partial_r \dot{e}_g$, as described by Equation (\ref{eq:dedr}) but scaling out the $(1-e^2)/e$ trend to highlight the role of the torque and power densities. We first remark that from the plots of the integrated contribution to binary eccentricity as a function of radius plotted in the bottom row of Figure \ref{fig:eccDiskProf}, the innermost parts of the accretion flow (at $r<a$) contribute relatively little to the eccentricity of the binary at low eccentricities, but act to damp the eccentricity of the binary at higher eccentricities. Comparing the profiles of $\langle\Sigma\rangle$ and $\partial_r \dot{e}_g$, the most significant contribution to the eccentric forcing of each binary comes from the inner edge of the circumbinary disk and within the cavity. As illustrated in Figure \ref{fig:surface}, the streams of material accreting onto and being flung away from the binary must then play a crucial its evolution.

Similarly to the accretion of thin disks onto circular binaries \citep{2022MNRAS.513.6158D,2025ApJ...984..144T}, a smaller fraction of each accretion stream is captured by the binary each time one of its members passes by the pericenter of the circumbinary disk at higher Mach numbers (as pressure gradients become less significant), causing more to be flung away into the cavity edge, carrying off energy and angular momentum from the binary per unit accreted mass. When $\dot{J}/\dot{M}<0$, this results in a buildup of material in the inner regions of the disk as its structure approaches a steady state \citep[e.g.,][]{2013ApJ...774..144R,2022MNRAS.513.6158D,2025ApJ...984..144T}. This trend explains the magnitude of $\partial_r \dot{e}_g$ with Mach number, but not how its sign changes (for example, how $\dot{e}$ becomes negative by $e=0.45$ at $\mathcal{M}$, by $e=0.5$ at $\mathcal{M}=20$, but is still positive at $e=0.6$ at $\mathcal{M}=30$). From Figure \ref{fig:surface}, we see that the stronger shocks and weaker pressure gradients of high-Mach-number disks result in more evacuated cavities and thinner streams flowing to and from the binary; both factors increase the asymmetry of the accretion flow and tend towards sapping proportionally more angular momentum from the binary, as we will see in more detail in the following section.

\subsubsection{Attractors at High Eccentricities or the Lack Thereof }\label{sec:attractors}
Previous studies have identified stable binary eccentricities at a range of values depending on the disk properties: for example, when studying equal-mass binaries accreting from $\mathcal{M}=10$ disks, \citet{2021ApJ...909L..13Z} found that the binary eccentricity $e=0.445$ would be stable, while \citet{2021ApJ...914L..21D} found that $e\approx0.39$ would be stable, and later \citet{2023MNRAS.522.2707S} suggested a steady state around $0.4<e<0.5$. When studying unequal-mass binaries accreting from massive, self-gravitating disks, using particle-based codes,  \citet{2011MNRAS.415.3033R} inferred a steady steady-state eccentricity $0.6\lesssim e\lesssim 0.8$ and \citet{2024A&A...688A.174F} found a variety of steady-state configurations from $e\sim0.39$ to $e\sim0.9$ depending on the binary mass ratio and disk cooling rate. \citet{2025MNRAS.537.2422P} also identified numerous potential steady-state eccentricities for binaries with mass ratio $0.29$ for disks over a range of Mach numbers $10<\mathcal{M}<33$ and alpha viscosities $10^{-4}\leq\alpha\leq10^{-2}$, though that work only surveyed binary eccentricities as high as $e=0.4$.

Historical models of binary-disk interactions have focused predominantly on resonances, thanks to their \textit{necessity} in linear theory and the analytical tractability thereof; such models are typically valid when $q\ll\mathcal{M}^{-3}$, although they may also be valid in the circumbinary disk proper, well outside of the cavity. If only outer Lindblad resonances operate, without the potentially countervailing effects of corotation resonances or eccentric inner Lindblad resonances, they would excite binary eccentricities inexorably towards $e\sim1$ \citep{1980ApJ...241..425G,1991ApJ...370L..35A}. However, once binaries reach sufficiently high eccentricities, they can couple with the circumbinary disk through a variety of eccentricity-damping resonances, which can drive the total rate of change of the binary towards zero at $e\sim0.5-0.7$ \citep{1992btsf.work..145L,2000prpl.conf..731L}; one should keep in mind that the preceding resonance-based arguments predict no dependence at all of $\dot{a}/a$ or $\dot{e}$ on $\mathcal{M}$ at constant $\nu$. However, given that streams within the cavity region play a crucial role in determining the orbital evolution of the binary and their sensitivity to disk thermodynamics \citep{2020ApJ...900...43T,2022MNRAS.513.6158D,2025ApJ...984..144T}, it is unsurprising that these qualitative resonance-based arguments cannot explain our results in detail. Indeed, from Figure \ref{fig:eccDiskProf} we see that a substantial contribution to the change in eccentricity of the binary comes from within the cavity, where the flow highly variable and nearly ballistic.

\begin{figure}
\includegraphics[width=\linewidth]{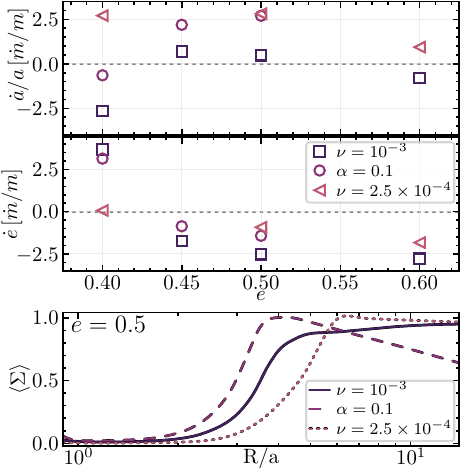}
\caption{The rates of change of binary semi-major axes and eccentricities for different disk viscosities, holding $\mathcal{M}=10$ constant. The bottom panel illustrates surface densities from the $e=0.5$ simulations for reference. Notably, the $\nu=2.5\times10^{-4}$ disks have much larger cavities than those with $\nu=10^{-3}$, resulting not in higher equilibrium eccentricities but instead mores subdued orbital evolution and possibly \textit{lower} equilibrium eccentricities. Moreover, disks with $\alpha$ viscosities have slightly smaller cavities, yet drive binaries to somewhat larger equilibrium eccentricities.}
\label{fig:orbitalVisc}
\end{figure}

Moving away from mechanisms inspired by linear theory, one suggestion for the origin of stable eccentricities was put forward by \citet{2011MNRAS.415.3033R}: that whether binary eccentricities are damped or excited depends on the relative angular velocity between the binary at apocenter and the disk at pericenter. This predicts that smaller cavities should be associated with lower equilibrium eccentricities --- a general trend borne out by the mass ratio survey carried out by \citet{2023MNRAS.522.2707S}. To test this hypothesis, we conducted a small suite of simulations at $0.4\leq e \leq 0.6$ using an alpha viscosity (specifically $\alpha=0.1$) to assess the impact of the functional form of the viscosity, and a smaller constant kinematic viscosity of $\nu=2.5\times10^{-4}$, which resulted in significantly larger cavities compared to our standard viscosity of $\nu=10^{-3}$. The results of these simulations are displayed in Figure \ref{fig:orbitalVisc}. Despite having much larger cavities, the low-viscosity simulations would predict, if anything, lower equilibrium eccentricities rather than higher.\footnote{In passing, we mention the suggestion made in \citet{2021ApJ...914L..21D} that the sign of $\dot{e}$ might be related to whether or not the circumbinary disk precesses or is locked in place relative to the orientation of the binary; similar to \citet{2023MNRAS.522.2707S}, we find no support for this as the disks precess similarly on both sides of the attractor.}

\begin{figure}
\includegraphics[width=\linewidth]{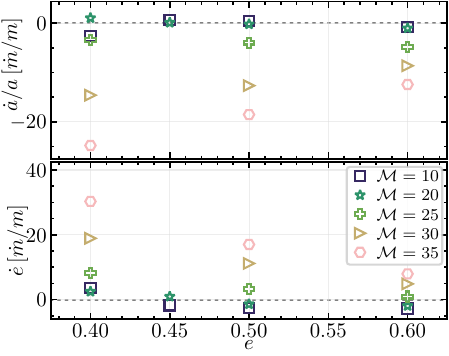}
\caption{The rates of change of binary semi-major axes and eccentricities for different disk Mach numbers, holding $\nu=10^{-3}$ constant. Rates of eccentricity excitation and putative equilibrium eccentricities appear to increase monotonically for $\mathcal{M}\geq20$. }
\label{fig:orbitalMach}
\end{figure}

\begin{figure*}
\includegraphics[width=\linewidth]{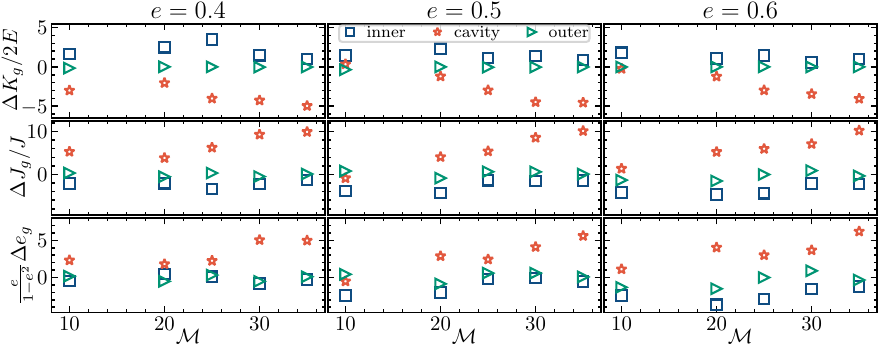}
\caption{The change in binary energy, angular momentum, and eccentricity due to interactions with various regions of the disk. Blue squares plot contributions from the vicinity of the minidisks, orange stars plot contributions from within the cavity, and green triangles plot contributions from the circumbinary disk itself (see the final two paragraphs of Section \ref{sec:attractors} for details).}
\label{fig:dedrMach}
\end{figure*}

Although the preceding discussion and Figure \ref{fig:orbitalScatter} illustrated the dramatic changes in binary orbital evolution when accreting from a thin disk, following from the reduced importance of pressure gradients and increased shock strengths driving dissipation, the coarse sampling in Mach number leaves some room for doubt as to the generality of the trend. To complement our main suite of simulations, we have also conducted a small survey of simulations at $\mathcal{M}=25$ and $\mathcal{M}=35$, which are presented in Figure \ref{fig:orbitalMach}. These confirm that the transition between our $\mathcal{M}=20$ and $\mathcal{M}=30$ simulations was not discontinuous, but part of a smooth trend; apart from the transition between $\mathcal{M}=10$ and $\mathcal{M}=20$, trends in $\dot{a}$ and $\dot{e}$ are monotonic with Mach number. Barring some other change in the nature of the accretion flow at very high Mach numbers, these results suggest that binaries accreting from thinner disks should be driven to increasingly high eccentricities at rates orders of magnitude higher than binaries accreting from thicker $h/r\sim0.1$ disks. 

The balance between the capture and repulsion of gaseous streams from the binary, which is responsible in the circular equal-mass case for determining the inspiral or outspiral of binaries and the rate thereof \citep{2022MNRAS.513.6158D,2025ApJ...984..144T}, can qualitatively explain the \textit{magnitude} of $\dot{e}$, but not the shift in the high-$e$ fixed point towards higher eccentricities. 
As pressure support in the fluid is reduced, the streams themselves become thinner and the gas shocks more strongly. The former might change the balance between the energy and angular momentum exchanged between the binary and disk to shift, and the latter might affect the inner edge of the circumbinary disk and the minidisks themselves.

To investigate these effects, we have integrated the gravitational contributions to the binary angular momentum, energy, and eccentricity over different regions of the accretion flow (the innermost regions, where most gas is bound to one of the two objects; the high-density circumbinary disk; and the intermediate cavity region)
as functions of Mach number for a few binary eccentricities; the results of this exercise are shown in Figure \ref{fig:dedrMach}. These quantities are computed by integrating Equations (\ref{eq:djdr}), (\ref{eq:dkdr}) and (\ref{eq:dedr}) respectively, and are scaled according to their prefactors in Equation (\ref{eq:dedr}). For each quantity, the `inner' contribution is calculated by integrating from $r=0$ to $r=0.5(1+e)+0.37$, roughly the extent of the Roche lobe of the binary at apocenter; the `outer' contribution integrates over the region of the disk that is not strongly perturbed by stream dynamics, and is defined by where the mass-weighted averaged fluid semi-major axis within an annulus matches the radius of that annulus to better than 2\%; the `cavity' contribution integrates over the intermediate region. We caution that this division between contributions within the cavity and owing to the circumbinary disk proper is imprecise, while the division between contributions from the innermost and other regions is fairly precise.

Figure \ref{fig:dedrMach} illustrates that while the contributions from the minidisks (`inner') and circumbinary disk (`outer') have some scatter as $e$ and $\mathcal{M}$ vary, they do not depend systematically on the disk thickness. On the other hand, the contribution from the streams of gas within the cavity (`cavity') depends strongly on the disk Mach number: the thinness of the streams, and possibly the strength of shocks against the inner edge of the circumbinary disk, are evidently responsible for the shift of the fixed-point in binary eccentricity moving upwards with disk Mach number. We also note that eccentricity damping from the mindisks becomes much more important at $e=0.5$ and $e=0.6$ than at $e=0.4$, as the closer pericenter approaches lead to more shocks and other dissipation which sap energy from the binary without much affecting its angular momentum (from the high velocities and short lever arms of the binary orbit at that point). It is this pericentric damping which eccentricity excitation from the streams must overcome to drive the binary eccentricity upward.

\subsubsection{The Stability of Nearly-Circular Binaries}\label{sec:nearcirc}

\begin{figure}
\includegraphics[width=\linewidth]{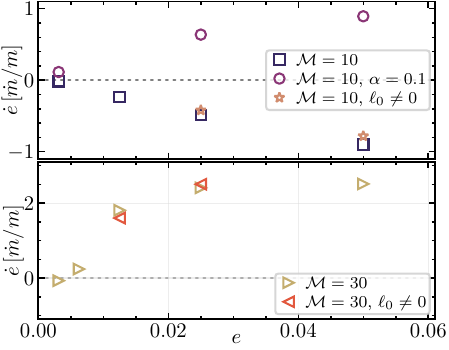}
\caption{The rate of change of binary eccentricity for nearly-circular binaries: the top panel plots the results for $\mathcal{M}=10$ disks, and the bottom for $\mathcal{M}=30$ disks. Squares in the top panel and right-pointing triangles in the bottom panel plot the results using our fiducial viscosity (constant-$\nu$) and $\ell_0=0$ in Equation \ref{eq:initSig}. Stars in the top panel and left-pointing triangles plot results with $\ell$ chosen to initiate each disk with its quasi-steady surface density profile (the value of $\dot{J}/\dot{M}$ measured from an initial $\ell=0$ run, see Appendix \ref{app:theTable}); whether or not each disk reaches a quasi-steady state is effectively irrelevant for $\dot{e}$ near $e\sim0$, and seems to have no bearing on whether or not near-circular binaries are driven towards higher or lower eccentricities. In the top panel, circles plot the results from a pair of simulations using an $\alpha=0.1$ viscosity; at $\mathcal{M}=10$ then, the viscosity prescription employed appears to determine the sign of $\dot{e}$ at low-but-nonzero binary eccentricities. Unlike thicker disks, even with constant-$\nu$ viscosities thin disks drive near-circular binaries to higher eccentricities.  }
\label{fig:orbitalStability}
\end{figure}

Whether or not near-circular orbits are circularized by interactions with their accretion disks is crucial for interpreting and modeling stellar and black hole binary systems. However, there has been substantial disagreement in the literature over whether or not near-circular binaries should be driven towards circularity or spurred on towards greater eccentricities, even when simulating nominally identical disks. For example, while simulating $\mathcal{M}=10$ disks with $\alpha=0.1$ viscosities, \citet{2021ApJ...909L..13Z} found that near-circular binaries were circularized, up to $e\lesssim0.08$, while \citet{2023MNRAS.522.2707S} found that all binaries with $e\lesssim0.45$ were driven towards higher eccentricities. When simulating $\mathcal{M}=10$ disks, but with a constant kinematic viscosity of $\nu=0.001$, \citet{2021ApJ...914L..21D} found that binaries with eccentricities $\lesssim0.06$ would be driven towards circularity. Reassuringly, our results at $\mathcal{M}=10$ and $\nu=0.001$, shown in Figure \ref{fig:orbitalScatter}, generally agree with the analogous simulations of \citet{2021ApJ...914L..21D}, suggesting that binaries should transition away from circularization somewhere between $e=0.05$ and $e=0.1$. However, at the same viscosity, we find that only binaries with $e\lesssim0.025$ are circularized at $\mathcal{M}=20$, and that all binaries with $e\in[0.0125,0.6]$ are driven towards yet higher eccentricities at $\mathcal{M}=30$. 

To verify the robustness of our aforementioned results on the stability of near-circular orbits, and in an attempt to shed light on disagreements between various earlier works at $\mathcal{M}=10$, we have conduced a handful of low-eccentricity simulations at $\mathcal{M}=10$ and $\mathcal{M}=30$ to test importance of differences in viscosity prescription and the importance of reaching a steady state, the results of which are presented in Figure \ref{fig:orbitalStability}. Besides the discrepancy of \citet{2023MNRAS.522.2707S} with \citet{2021ApJ...909L..13Z} and \citet{2021ApJ...914L..21D}, one might expect $\alpha$-disks to affect binaries differently from those with constant kinematic viscosity through the different density profiles they bring about in the inner regions of the disk (as illustrated in Figure \ref{fig:orbitalVisc}). All of the simulations contributing to Figure \ref{fig:orbitalScatter} were initialized setting $\ell_0=0$: when this value is below the eventual $\dot{J}/\dot{m}$ driven by the binary, accretion is temporarily enhanced as the inner regions of the accretion disk are denser than their eventual steady state, and when $\ell_0>\dot{J}/\dot{m}$ the accretion rate is temporarily reduced as the inner regions of the disk are less dense than their eventual steady state \citep{2017MNRAS.466.1170M,2022MNRAS.513.6158D}. Putting aside the question of whether or not one should expect astrophysical circumbinary disks to have enough time to reach these steady-state profiles, it is pertinent to determine the effects of these transient disk structures, and of initial conditions generally, on the stability of near-circular binaries. 

Turning to the effects of disk thickness while holding viscosity fixed, the neighborhood of stability for nearly-circular binaries (the range of near-zero eccentricities for which $\dot{e}<0$) is much larger in thick disks than thin: $\mathcal{M}=30$ disks seem to drive the eccentricities of binaries upward for $e\gtrsim0.005$; in the analogous $\mathcal{M}=10$ disks, the transition from circularization to eccentricity excitation lies closer to $e\approx0.07$. These appear quite insensitive to how far each simulation was initialized from a steady state: the values of $\dot{e}$ found in simulations initialized using nearly steady-state values of $\ell_0$ are nearly identical to those from simulations initialized using $\ell_0=0$, despite the latter simulations being far from a viscous steady state. 

For thicker ($\mathcal{M}=10$) disks, we find that disks with $\alpha=0.1$ viscosities cause near-circular binaries to become more eccentric, whereas those with constant-$\nu$ viscosities tend to damp binary eccentricities (for $e\leq0.05$, mind you). This is reminiscent of how differences between constant-$\alpha$ vs constant-$\nu$ viscosities strongly affect the balance of accretion between the constituents of unequal-mass binaries \citep{2024ApJ...967...12D}. Since constant-$\alpha$ vs constant-$\nu$ disks result in fairly different density profiles, particularly within the cavity and in the inner regions of the disk (Figure \ref{fig:orbitalVisc}), it is perhaps unsurprising that the resulting rate of change of the binary eccentricity should not also change. While this result seems to qualitatively agree with \citet{2023MNRAS.522.2707S}, it is unclear why \citet{2021ApJ...909L..13Z} found a discrepant result.

\section{Discussion}\label{sec:discussion}

\subsection{Supermassive Black Hole Binaries} 
As we have seen above, thin circumbinary disks can drive binaries to fairly high eccentricities. As binary black holes approach merger, gravitational waves will efficiently circularize their orbits; but at leading order the rates of change in the semi-major axis and eccentricity of the binary due to gravitational wave emission scale as $\dot{a}_{\rm GW}\propto a^{-3}$ and $\dot{e}_{\rm GW}\propto a^{-4}$ \citep{1964PhRv..136.1224P}, and thus at larger separations astrophysical eccentricity pumping will handily dominate. The question then becomes at what separations the character of the binary shifts from that sculpted by its astrophysical environment to one fixed by gravitational wave emission.

\begin{figure}
\includegraphics[width=\linewidth]{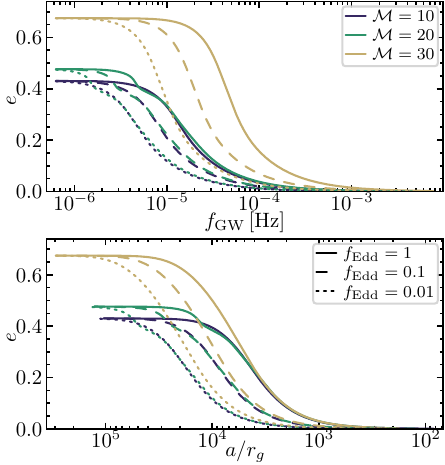}
    \caption{Illustrative tracks of SMBH binary eccentricity as a function of semi-major axis and the roughly millihertz gravitational wave frequencies (plotted here without cosmological redshift) that will be probed by LISA and other space-based gravitational wave interferometers. As an example, these calculations assume binaries with $m=10^5\,M_\odot$. These calculations assume accretion at various fraction of the Eddington-rate (with mass $e$-folding timescales of $5\times10^{7-9}$ years) and the corresponding changes in orbital evolution estimated in Section \ref{sec:orbital} due to accretion disks of various thicknesses, as well as the leading-order effects of gravitational wave emission \citep{1964PhRv..136.1224P}. Binaries will enter the LISA band with high eccentricities (to an extend that depends on disk thickness) and gradually circularize as they approach merger.}\label{fig:lisaGW}
\end{figure}

\begin{figure}
    \includegraphics[width=\linewidth]{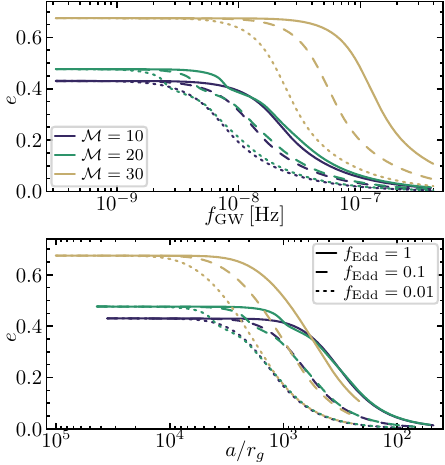}
    \caption{The same exercise as shown in Figure \ref{fig:lisaGW} but for higher-mass ($m=10^9\,M_\odot$) SMBH binaries with lower gravitational wave frequencies relevant to pulsar timing arrays. Higher-mass binaries will remain highly eccentric until fractionally closer separations. }\label{fig:ptaGW}
\end{figure}

To gain a sense for the typical eccentricities we might expect binary supermassive black holes to possess as they emit low-frequency gravitational waves, we have conducted a small set of illustrative orbital evolution calculations: given an disk Mach number and binary orbital parameters (initial mass, semi-major axis, and eccentricity), we have integrated those orbital parameters until merger accounting for gravitational wave emission at leading order \citep{1964PhRv..136.1224P} and fitting interpolating cubic splines to the orbital evolution results shown in Figure \ref{fig:orbitalScatter}, for a few different accretion rates, as fractions $f_{\rm Edd}$ of the Eddington rate. We targeted one set of calculations (Figure \ref{fig:lisaGW}) towards millihertz-frequency gravitational wave detectors such as LISA, Taiji, and TianQin \citep[e.g.,][]{2021NatAs...5..881G,2023LRR....26....2A} and another towards pulsar timing arrays (Figure \ref{fig:ptaGW}) by selecting particular masses and initial separations. We estimate the characteristic gravitational wave frequency for eccentric binaries using the fit provided in \citet{2021RNAAS...5..275H}.\footnote{As binaries approach merger, this notion of eccentricity is imprecise, but it suffices well before merger \citep[see, e.g.,][for discussion of this issue]{2024ApJ...969..132V}.}

Generally speaking, we expect binaries to enter the PTA and millihertz gravitational wave bands with substantial eccentricities, $e\gtrsim0.6$ for thin disks, and for those binaries to circularize. Higher-mass binaries that emit lower-frequency gravitational waves, if actively accreting, may circularize fairly late in their inspirals, at separations less than $\sim100 r_g$, where $r_g\equiv Gmc^{-2}$; the lower-mass binaries that emit gravitational waves in the millihertz band will circularize later in their inspirals.\footnote{In this case ``later'' means at smaller values of $a/r_g$. This result follows trivially from the scalings of $\dot{e}_{\rm GW}$ and $\dot{a}_{\rm GW}$ with binary separation quoted above.} Binaries can therefore maintain fairly high eccentricities well into the pulsar timing array band, while we may observe the circularization of binaries at millihertz frequencies in real time. At higher accretion rates, the influence of the circumbinary disk can naturally overcome damping due to gravitational waves until later in the inspiral (the orbital frequency at which gravitational waves overcome the disk scales roughly as $\propto\dot{m}^{3/8}$). This dependence is fairly weak, and thus given the typical scatter in AGN accretion rates \citep[e.g.,][]{2008ApJ...680..169S}, any accreting SMBH binary should maintain high eccentricities into the nanoHertz and milliHertz bands.  

The potential signatures of binary black holes, in both electromagnetic radiation and gravitational wave emission, depend strongly on binary eccentricity. Qualitatively, accretion onto near-circular binaries is often strongly on the orbital period of fluid near the inner edge of the circumbinary disk, typically $\sim4-5$ binary orbital periods; accretion onto eccentric binaries, on the other hand, tends to pulse strongly on the orbital period of the binary \citep[e.g.,][]{2021ApJ...909L..13Z,2022PhRvD.106j3010W}. Signatures of Doppler boosting will of course also evolve with both the separation and eccentricity of each binary \citep[e.g.,][]{2024ApJ...977..244D}. Somewhat provocatively, binary eccentricity inferred for long-term broad line radial velocity shifts in the quasar J0950+5128 is $e\sim0.65$ \citep{2025arXiv250506221M}, near the expected equilibrium value for binaries accreting from thin disks. 

Any electromagnetic emission from inspiralling black hole binaries should change in character, potentially quite significantly, as the binaries circularize. Circularization should occur while binaries are still well-coupled and actively accreting from their disks, which will break down when $\dot{a}_{\rm GW}$ exceeds the rate at which the inner edge of the disk can spread \citep[e.g.,][]{2002ApJ...567L...9A,2023ApJ...949L..30D}; potential signatures of circularization should thus be distinct from those of the decoupling process \citep[e.g.,][]{2023ApJ...949L..30D,2023MNRAS.526.5441K,2025MNRAS.537.3620Z,2025arXiv250206389E}. 

Gravitational wave emission is of course strongly peaked about pericentre, and as binaries inspiral through the millihertz band changes in their gravitational waveforms should be apparent. Because thin disks excite binary eccentricity so rapidly, they may measurably alter binary inspirals; because disk-driven eccentricity excitation depends so sensitively on disk thickness, as illustrated in Figures \ref{fig:lisaGW} and \ref{fig:ptaGW}, gravitational waves alone might provide (at least broad and qualitative) constraints on AGN disk structure. For unresolved sources contributing the to stochastic gravitational wave background probed by pulsar timing arrays, the crucial difference in gravitational wave emission is that it occurs not at a single frequency as for quasi-circular binaries, but over many harmonics. This results in an eccentricity-dependent turnovers and excesses in the stochastic GW power spectrum at frequencies of $\sim10^{-9}-10^{-8}\,{\rm Hz}$ for binary populations with $e\sim0.6-0.9$ \citep[e.g.][]{2017MNRAS.470.1738C}, which pulsar timing arrays should be able to constrain given longer observing baselines \citep[c.f.,][]{2023ApJ...951L...8A,2023A&A...678A..50E}. Furthermore, the dispersal of gravitational wave emission between harmonics complicates the search for anisotropy in the stochastic GW background within a given frequency bin \citep[e.g.,][]{2024PhRvD.109l3544S}.

\subsection{Stellar Binaries}
Circumbinary accretion disks play can play a significant role in shaping the architectures of stellar binaries, both during their formation \citep[e.g.,][]{1986ApJS...62..519B,1997MNRAS.285...33B,2023MNRAS.521.5334P} and in the later stages of their lives as stars shed their envelopes \citep[e.g.,][]{1998A&A...332..877J,2018A&A...620A..85O}, although the potential effects of disks on stellar binaries can be limited by the mass available and time between their formation and observation. In particular, many post-AGB binaries are thought to have been tidally circularized during earlier stages of their evolution, followed by disk-driven eccentricity excitation \citep[e.g.,][]{2013A&A...551A..50D}. The bimodality of post-AGB binary eccentricities \citep[e.g.,][]{2018A&A...620A..85O}, many nearly-circular and others with $e$ as high as $\sim0.6$, can result from differences in the degree of tidal circularization and the state of the later-forming accretion disks: if binaries are only driven down to $e\sim0.06$ or so, then they might be subsequently driven to $e\sim0.4-0.7$ depending on the thickness of the resulting disk. On the other hand, whether or not binaries with $e\sim0$ can be circularized depends sensitively on both the disk thickness and viscosity, as discussed in Section \ref{sec:nearcirc}.

In the context of star formation, the combination of inward migration and eccentricity excitation driven by thin disks may be related to the high eccentricities of many short-period F-star binaries \citep[e.g.,][]{2024ApJ...975..149M}. Closer-separation $(a\lesssim200~\rm{AU})$ binaries seem to have sub-thermal eccentricity distributions \citep{2020MNRAS.496..987T}, consistent with disk-driven eccentricity damping at $e\gtrsim0.4-0.7$ depending on the disk thickness.\footnote{The eccentricity distributions of close binaries often seem better described by Rayleigh distributions than by power laws \citep{2025ApJ...982L..34W}, so one must take care when interpreting inferences of binary eccentricities relative to a thermal distribution.} 
On the other hand, this makes disk-driven migration unlikely to be the sole cause for wide-separation eccentric nearly-equal-mass (``twin'') binaries \citep[e.g.,][]{2022ApJ...933L..32H}; while thin circumbinary disks may help equalize binary mass ratios \citep[e.g.][]{2015MNRAS.447.2907Y},\footnote{One should keep in mind that the result that accretion rates are higher onto the lower-mass component of a binary only holds in \textit{thin} disks, while thicker disk can accrete primarily onto the primary \citep{2015MNRAS.447.2907Y}. See, for example, \citet{2005ApJ...623..922O} and \citet{2010ApJ...708..485H} for early examples where this was observed, and \citet{Dittmann:2025aup} for an example in the more extreme case of accretion from disks with $H>a$.} the disks capable of driving binaries to eccentricities $e\gtrsim0.9$ (rather than damping eccentricities above $e\gtrsim0.4$) would almost certainly drive binaries to smaller separations, and thus the orbital elements of highly eccentric (superthermally distributed) wide binaries are likely to be sculpted by other factors such as turbulence in the interstellar medium \citep[e.g.,][]{2023ApJ...949L..28X}.

\subsection{Caveats}
We expect our result that thin disks drive binaries to high eccentricities to be qualitatively robust. However, the precise rates at which binary orbital elements change will be strongly affected by physics we have neglected. Furthermore, the breadth of our simulations was limited, and extension to other mass ratios, higher Mach numbers, and lower viscosities may be desirable.

Our analysis in Section \ref{sec:orbital} illustrated that the streams of material stripped from the inner disk and then flung away from the binary play a crucial role in the evolution of the binary orbit, mediating the exchange of energy and angular momentum between the binary and the disk. These dynamics are dominated by the tidal gravity of the binary and are quite generic, from simple Newtonian calculations like the ones presented here to more encompassing simulations including disk self-gravity, magnetic fields, radiation, and/or general relativity \citep[e.g.,][]{2012ApJ...749..118S,2021ApJ...922..175N,2024MNRAS.534.3448B,2025ApJ...986..158T}. Even the most highly magnetized AGN accretion disks seem to have midplanes with roughly equal magnetic and thermal pressure support \citep{2025OJAp....8E..39S,2025arXiv250512671G}, suggesting that the binary dynamics they drive might be similar to disks with dynamically subdominant magnetic fields, though this needs to be confirmed. However, particularly if the disk is threaded by a net vertical magnetic flux, binary evolution might be very strongly affected; under such conditions magnetically driven winds can carry off significant angular momentum from the accretion flow, spurring even faster orbital evolution \citep{2024ApJ...973L..19M,2025arXiv250816855W}. It also seems plausible that other treatments of disk thermodynamics or cooling would result in quantitatively different rates of binary orbital evolution, but unlikely that general trends would change, based on the few studies carried out for $\mathcal{M}\sim10$ disks \citep[e.g.,][]{2023MNRAS.526.3570W,2025arXiv250816855W}.

It seems probable that equilibrium binary eccentricities, and the rates of eccentricity excitation will decrease with mass ratio $q=m_2/m_1$, based on previous simulations of $\mathcal{M}=10$ disks down to $q=0.1$ \citep{2023MNRAS.522.2707S}. For circular binaries, similar trends in orbital evolution rate seem to hold until even lower mass ratios at $\mathcal{M}=10$, but the evolution of binaries in thinner disks can change dramatically at $q\lesssim0.2$, in some cases leading to orders-of-magnitude faster orbital evolution than for equal-mass binaries \citep{2024ApJ...967...12D}. The qualitative result that thinner disks drive binaries more quickly to higher eccentricities seems to hold at $q=0.26$ \citep{2025MNRAS.537.2422P}, although the generality of our results at lower mass ratios is a major uncertainty. 

One potential limitation of our survey thus far is that we initialized most simulations with $\ell_0=0$ in Equation (\ref{eq:initSig}), while many binaries accrete with $|\dot{J}/\dot{m}| \gg 1$, as shown by Figure \ref{fig:gravitationalScatter}; crucially, this means that binaries with $\dot{J}/\dot{m}>0$ tend to accrete at a higher-than-steady-state rate, while those with $\dot{J}/\dot{m}<0$ tend to accrete below the steady state rate while the disk structure adjusts \citep{2017MNRAS.466.1170M,2022MNRAS.513.6158D}. This adjustment typically results in much smaller changes to $\dot{J}/\dot{m}$ than to $\dot{m}$ itself, though this has been tested most thoroughly for circular binaries. Figure \ref{fig:orbitalStability} illustrates that $\dot{e}$ is negligibly affected by simulations starting out of equilibrium for nearly circular binaries. Appendix \ref{app:theTable} also records the results of a couple tests at $e=0.5$ and $e=0.6$ for $\mathcal{M}=30$ disks, which show that those results too are very weakly sensitive to whether or not the disk is initialized near a quasi-steady state. 
One should also keep in mind that the astrophysical timescales required for a system to reach a steady state might make the notion irrelevant: as shown by \citet{2025ApJ...984..144T}, realistically thin AGN disks ($\mathcal{M}\gtrsim 100$) may take millions (or more) viscous timescales to reach a quasi-steady state, potentially longer than a Hubble time.

\section{Summary}\label{sec:summary}
We have conducted a broad survey of circumbinary disk simulation up to binary eccentricities of $0.6$, surveying different disk thicknesses and disk viscosity models. Thin disks drive binaries to higher eccentricities (at higher rates) compared to thin disks; thin disks also drive binaries to inspiral regardless of their eccentricity, whereas thicker disks can cause binaries to outspiral. The stability of near-circular binaries, whether a low-eccentricity binary will be damped towards $e=0$ or driven towards some higher eccentricity, depends sensitively on disk thickness and viscosity: at $\mathcal{M}=10$, disks with a constant kinematic viscosity ($\nu$) damp binaries towards circularity and disks with a constant-$\alpha$ viscosity drive binary eccentricities upward; thin disks, even modeled with a constant kinematic viscosity, spur binary eccentricities upwards above $e\sim0.003$. 

Our results suggest that many binary supermassive black holes, which are often thought to accrete from thin disks, should have eccentricities in excess of $e\gtrsim0.6$ until rather close to merger, at roughly $a\gtrsim10^2-10^3\,r_g$. Most LISA sources should gradually circularize throughout their inspirals, from initial eccentricities related to their environments. The stochastic gravitational wave spectrum emitted from a population of eccentric supermassive black hole binaries is suppressed at low-frequencies and nearly identical at high frequencies relative to a population of circular binaries, often with exceeds strain around $\sim10^{-9}$ Hz \citep[e.g.,][]{2017MNRAS.470.1738C}. Furthermore, this redistribution of strain to higher harmonics will affect searches for anisotropy \citep[e.g.,][]{2023ApJ...956L...3A} within a given bandwidth. Our results seem inconsistent with observations of wide stellar binaries, which display superthermal eccentricities \citep[e.g.][]{2022ApJ...933L..32H}, since disks able to produce such eccentricity pumping would shrink rather than expand orbits, suggesting other mechanisms to be responsible \citep[e.g.,][]{2023ApJ...949L..28X}; however, disks may damp the upper end of thermally distributed eccentricities, helping to produce the observed subthermal eccentricities observed in closer binaries \citep[e.g.,][]{2020MNRAS.496..987T}.

\section*{Software}
\texttt{Disco} \citep{2016ApJS..226....2D,2021ApJ...921...71D}, \texttt{matplotlib} \citep{4160265}, \texttt{cmocean} \citep{cmocean}, \texttt{numpy} \citep{5725236}

\section*{Acknowledgments}
AJD thanks Nadia Zakamska, Steve Lubow, Yanqin Wu, and Yoram Lithwick for insightful conversations over the course of this project.
Support for this work was provided by NASA
through the NASA Hubble Fellowship grant No. HST-HF2-
51553.001, awarded by the Space Telescope Science Institute,
which is operated by the Association of Universities for Research
in Astronomy, Inc., for NASA, under contract NAS5-26555.
L.C.~is a CITA National fellow and acknowledges the support by the Natural Sciences and Engineering Research Council of Canada (NSERC), funding reference DIS-2022-568580.
Research at Perimeter Institute is supported in part by the Government of Canada through the Department of Innovation, Science and Economic Development and by the Province of Ontario through the Ministry of Colleges and Universities.
This work was performed in part at Aspen Center for Physics, which is supported by National Science Foundation grant PHY-2210452.
\pagebreak

\bibliographystyle{aasjournal}
\bibliography{references}

@ARTICLE{2016ApJS..226....2D,
       author = {{Duffell}, Paul C.},
        title = "{DISCO: A 3D Moving-mesh Magnetohydrodynamics Code Designed for the Study of Astrophysical Disks}",
      journal = {\apjs},
         year = 2016,
        month = sep,
       volume = {226},
       number = {1},
          eid = {2},
        pages = {2},
          doi = {10.3847/0067-0049/226/1/2},
archivePrefix = {arXiv},
       eprint = {1605.03577},
 primaryClass = {physics.comp-ph},
       adsurl = {https://ui.adsabs.harvard.edu/abs/2016ApJS..226....2D}
}

@ARTICLE{2008ApJ...686..432S,
       author = {{Sesana}, Alberto and {Haardt}, Francesco and {Madau}, Piero},
        title = "{Interaction of Massive Black Hole Binaries with Their Stellar Environment. III. Scattering of Bound Stars}",
      journal = {\apj},
         year = 2008,
        month = oct,
       volume = {686},
       number = {1},
        pages = {432-447},
          doi = {10.1086/590651},
archivePrefix = {arXiv},
       eprint = {0710.4301},
 primaryClass = {astro-ph},
       adsurl = {https://ui.adsabs.harvard.edu/abs/2008ApJ...686..432S}
}

@ARTICLE{2007A&A...463..683S,
       author = {{S{\"o}derhjelm}, S.},
        title = "{The q = 1 peak in the mass-ratios for Hipparcos visual binaries}",
      journal = {\aap},
         year = 2007,
        month = feb,
       volume = {463},
       number = {2},
        pages = {683-691},
          doi = {10.1051/0004-6361:20066024},
       adsurl = {https://ui.adsabs.harvard.edu/abs/2007A&A...463..683S}
}

@ARTICLE{2022ApJ...933L..32H,
       author = {{Hwang}, Hsiang-Chih and {El-Badry}, Kareem and {Rix}, Hans-Walter and {Hamilton}, Chris and {Ting}, Yuan-Sen and {Zakamska}, Nadia L.},
        title = "{Wide Twin Binaries are Extremely Eccentric: Evidence of Twin Binary Formation in Circumbinary Disks}",
      journal = {\apjl},
         year = 2022,
        month = jul,
       volume = {933},
       number = {2},
          eid = {L32},
        pages = {L32},
          doi = {10.3847/2041-8213/ac7c70},
archivePrefix = {arXiv},
       eprint = {2205.05690},
 primaryClass = {astro-ph.SR},
       adsurl = {https://ui.adsabs.harvard.edu/abs/2022ApJ...933L..32H}
}

@ARTICLE{2020MNRAS.496..987T,
       author = {{Tokovinin}, Andrei},
        title = "{Eccentricity distribution of wide low-mass binaries}",
      journal = {\mnras},
         year = 2020,
        month = jul,
       volume = {496},
       number = {1},
        pages = {987-993},
          doi = {10.1093/mnras/staa1639},
archivePrefix = {arXiv},
       eprint = {2004.06570},
 primaryClass = {astro-ph.SR},
       adsurl = {https://ui.adsabs.harvard.edu/abs/2020MNRAS.496..987T}
}

@ARTICLE{2023ApJ...949L..28X,
       author = {{Xu}, Siyao and {Hwang}, Hsiang-Chih and {Hamilton}, Chris and {Lai}, Dong},
        title = "{Wide-binary Stars Formed in the Turbulent Interstellar Medium}",
      journal = {\apjl},
         year = 2023,
        month = jun,
       volume = {949},
       number = {2},
          eid = {L28},
        pages = {L28},
          doi = {10.3847/2041-8213/acd6f7},
archivePrefix = {arXiv},
       eprint = {2303.16224},
 primaryClass = {astro-ph.GA},
       adsurl = {https://ui.adsabs.harvard.edu/abs/2023ApJ...949L..28X}
}

@ARTICLE{2020ApJ...900...43T,
       author = {{Tiede}, Christopher and {Zrake}, Jonathan and {MacFadyen}, Andrew and {Haiman}, Zoltan},
        title = "{Gas-driven Inspiral of Binaries in Thin Accretion Disks}",
      journal = {\apj},
         year = 2020,
        month = sep,
       volume = {900},
       number = {1},
          eid = {43},
        pages = {43},
          doi = {10.3847/1538-4357/aba432},
archivePrefix = {arXiv},
       eprint = {2005.09555},
 primaryClass = {astro-ph.GA},
       adsurl = {https://ui.adsabs.harvard.edu/abs/2020ApJ...900...43T}
}

@ARTICLE{2020ApJ...892L..29D,
       author = {{Dempsey}, Adam M. and {Mu{\~n}oz}, Diego and {Lithwick}, Yoram},
        title = "{Inner Boundary Condition in Quasi-Lagrangian Simulations of Accretion Disks}",
      journal = {\apjl},
         year = 2020,
        month = apr,
       volume = {892},
       number = {2},
          eid = {L29},
        pages = {L29},
          doi = {10.3847/2041-8213/ab800e},
archivePrefix = {arXiv},
       eprint = {2002.05164},
 primaryClass = {astro-ph.EP},
       adsurl = {https://ui.adsabs.harvard.edu/abs/2020ApJ...892L..29D}
}

@ARTICLE{2024ApJ...964...61D,
       author = {{Dittmann}, Alexander J. and {Dempsey}, Adam M. and {Li}, Hui},
        title = "{The Evolution of Inclined Binary Black Holes in the Disks of Active Galactic Nuclei}",
      journal = {\apj},
         year = 2024,
        month = mar,
       volume = {964},
       number = {1},
          eid = {61},
        pages = {61},
          doi = {10.3847/1538-4357/ad23ce},
archivePrefix = {arXiv},
       eprint = {2310.03832},
 primaryClass = {astro-ph.HE},
       adsurl = {https://ui.adsabs.harvard.edu/abs/2024ApJ...964...61D}
}

@article{Dittmann:2025aup,
    author = "Dittmann, Alexander J. and Dempsey, Adam M. and Li, Hui",
    title = "{The Multiple Paths to Merger of Unequal-mass Black Hole Binaries in the Disks of Active Galactic Nuclei}",
    eprint = "2505.05555",
    archivePrefix = "arXiv",
    primaryClass = "astro-ph.HE",
    doi = "10.3847/1538-4357/adea72",
    journal = "Astrophys. J.",
    volume = "990",
    number = "2",
    pages = "137",
    year = "2025"
}

@ARTICLE{2023MNRAS.526.3570W,
       author = {{Wang}, Hai-Yang and {Bai}, Xue-Ning and {Lai}, Dong and {Lin}, Douglas N.~C.},
        title = "{Hydrodynamical simulations of circumbinary accretion: balance between heating and cooling}",
      journal = {\mnras},
         year = 2023,
        month = dec,
       volume = {526},
       number = {3},
        pages = {3570-3588},
          doi = {10.1093/mnras/stad2884},
archivePrefix = {arXiv},
       eprint = {2212.07416},
 primaryClass = {astro-ph.HE},
       adsurl = {https://ui.adsabs.harvard.edu/abs/2023MNRAS.526.3570W}
}

@ARTICLE{4160265,
  author={J. D. {Hunter}},
  journal={Computing in Science   Engineering}, 
  title={Matplotlib: A 2D Graphics Environment}, 
  year={2007},
  volume={9},
  number={3},
  pages={90-95},
  doi={10.1109/MCSE.2007.55}}

@ARTICLE{5725236,
  author={S. {van der Walt} and S. C. {Colbert} and G. {Varoquaux}},
  journal={Computing in Science   Engineering}, 
  title={The NumPy Array: A Structure for Efficient Numerical Computation}, 
  year={2011},
  volume={13},
  number={2},
  pages={22-30},
  doi={10.1109/MCSE.2011.37}}

@ARTICLE{2020MNRAS.499.3362R,
       author = {{Ragusa}, Enrico and {Alexander}, Richard and {Calcino}, Josh and {Hirsh}, Kieran and {Price}, Daniel J.},
        title = "{The evolution of large cavities and disc eccentricity in circumbinary discs}",
      journal = {\mnras},
         year = 2020,
        month = dec,
       volume = {499},
       number = {3},
        pages = {3362-3380},
          doi = {10.1093/mnras/staa2954},
archivePrefix = {arXiv},
       eprint = {2009.10738},
 primaryClass = {astro-ph.EP},
       adsurl = {https://ui.adsabs.harvard.edu/abs/2020MNRAS.499.3362R}
}

@ARTICLE{2017MNRAS.466.1170M,
       author = {{Miranda}, Ryan and {Mu{\~n}oz}, Diego J. and {Lai}, Dong},
        title = "{Viscous hydrodynamics simulations of circumbinary accretion discs: variability, quasi-steady state and angular momentum transfer}",
      journal = {\mnras},
         year = 2017,
        month = apr,
       volume = {466},
       number = {1},
        pages = {1170-1191},
          doi = {10.1093/mnras/stw3189},
archivePrefix = {arXiv},
       eprint = {1610.07263},
 primaryClass = {astro-ph.SR},
       adsurl = {https://ui.adsabs.harvard.edu/abs/2017MNRAS.466.1170M}
}

@ARTICLE{1980ApJ...241..425G,
       author = {{Goldreich}, P. and {Tremaine}, S.},
        title = "{Disk-satellite interactions.}",
      journal = {\apj},
         year = 1980,
        month = oct,
       volume = {241},
        pages = {425-441},
          doi = {10.1086/158356},
       adsurl = {https://ui.adsabs.harvard.edu/abs/1980ApJ...241..425G}
}

@ARTICLE{1991ApJ...370L..35A,
       author = {{Artymowicz}, P. and {Clarke}, C.~J. and {Lubow}, S.~H. and {Pringle}, J.~E.},
        title = "{The Effect of an External Disk on the Orbital Elements of a Central Binary}",
      journal = {\apjl},
         year = 1991,
        month = mar,
       volume = {370},
        pages = {L35},
          doi = {10.1086/185971},
       adsurl = {https://ui.adsabs.harvard.edu/abs/1991ApJ...370L..35A}
}

@ARTICLE{2024A&A...688A.174F,
       author = {{Franchini}, Alessia and {Prato}, Alessandra and {Longarini}, Cristiano and {Sesana}, Alberto},
        title = "{The behaviour of eccentric sub-pc massive black hole binaries embedded in massive discs}",
      journal = {\aap},
         year = 2024,
        month = aug,
       volume = {688},
          eid = {A174},
        pages = {A174},
          doi = {10.1051/0004-6361/202449402},
archivePrefix = {arXiv},
       eprint = {2402.00938},
 primaryClass = {astro-ph.HE},
       adsurl = {https://ui.adsabs.harvard.edu/abs/2024A&A...688A.174F}
}

@ARTICLE{2025arXiv250512671G,
       author = {{Guo}, Minghao and {Quataert}, Eliot and {Squire}, Jonathan and {Hopkins}, Philip F. and {Stone}, James M.},
        title = "{Idealized Global Models of Accretion Disks with Strong Toroidal Magnetic Fields}",
      journal = {arXiv e-prints},
         year = 2025,
        month = may,
          eid = {arXiv:2505.12671},
        pages = {arXiv:2505.12671},
          doi = {10.48550/arXiv.2505.12671},
archivePrefix = {arXiv},
       eprint = {2505.12671},
 primaryClass = {astro-ph.HE},
       adsurl = {https://ui.adsabs.harvard.edu/abs/2025arXiv250512671G}
}

@ARTICLE{2025OJAp....8E..39S,
       author = {{Squire}, Jonathan and {Quataert}, Eliot and {Hopkins}, Philip F.},
        title = "{Rapid, strongly magnetized accretion in the zero-net-vertical-flux shearing box}",
      journal = {The Open Journal of Astrophysics},
         year = 2025,
        month = apr,
       volume = {8},
          eid = {39},
        pages = {39},
          doi = {10.33232/001c.136467},
archivePrefix = {arXiv},
       eprint = {2409.05467},
 primaryClass = {astro-ph.HE},
       adsurl = {https://ui.adsabs.harvard.edu/abs/2025OJAp....8E..39S}
}

@ARTICLE{2025MNRAS.537.2422P,
       author = {{Penzlin}, Anna B.~T. and {Booth}, Richard A. and {Nelson}, Richard P. and {Sch{\"a}fer}, Christoph M. and {Kley}, Wilhelm},
        title = "{Viscous circumbinary protoplanetary discs - II. Disc effects on the binary orbit}",
      journal = {\mnras},
         year = 2025,
        month = mar,
       volume = {537},
       number = {3},
        pages = {2422-2432},
          doi = {10.1093/mnras/staf177},
archivePrefix = {arXiv},
       eprint = {2501.17055},
 primaryClass = {astro-ph.EP},
       adsurl = {https://ui.adsabs.harvard.edu/abs/2025MNRAS.537.2422P}
}

@ARTICLE{2023MNRAS.522.2707S,
       author = {{Siwek}, Magdalena and {Weinberger}, Rainer and {Hernquist}, Lars},
        title = "{Orbital evolution of binaries in circumbinary discs}",
      journal = {\mnras},
         year = 2023,
        month = jun,
       volume = {522},
       number = {2},
        pages = {2707-2717},
          doi = {10.1093/mnras/stad1131},
archivePrefix = {arXiv},
       eprint = {2302.01785},
 primaryClass = {astro-ph.HE},
       adsurl = {https://ui.adsabs.harvard.edu/abs/2023MNRAS.522.2707S}
}

@ARTICLE{2011MNRAS.415.3033R,
       author = {{Roedig}, C. and {Dotti}, M. and {Sesana}, A. and {Cuadra}, J. and {Colpi}, M.},
        title = "{Limiting eccentricity of subparsec massive black hole binaries surrounded by self-gravitating gas discs}",
      journal = {\mnras},
         year = 2011,
        month = aug,
       volume = {415},
       number = {4},
        pages = {3033-3041},
          doi = {10.1111/j.1365-2966.2011.18927.x},
archivePrefix = {arXiv},
       eprint = {1104.3868},
 primaryClass = {astro-ph.CO},
       adsurl = {https://ui.adsabs.harvard.edu/abs/2011MNRAS.415.3033R}
}

@ARTICLE{2021ApJ...909L..13Z,
       author = {{Zrake}, Jonathan and {Tiede}, Christopher and {MacFadyen}, Andrew and {Haiman}, Zolt{\'a}n},
        title = "{Equilibrium Eccentricity of Accreting Binaries}",
      journal = {\apjl},
         year = 2021,
        month = mar,
       volume = {909},
       number = {1},
          eid = {L13},
        pages = {L13},
          doi = {10.3847/2041-8213/abdd1c},
archivePrefix = {arXiv},
       eprint = {2010.09707},
 primaryClass = {astro-ph.HE},
       adsurl = {https://ui.adsabs.harvard.edu/abs/2021ApJ...909L..13Z}
}

@ARTICLE{2021ApJ...914L..21D,
       author = {{D'Orazio}, Daniel J. and {Duffell}, Paul C.},
        title = "{Orbital Evolution of Equal-mass Eccentric Binaries due to a Gas Disk: Eccentric Inspirals and Circular Outspirals}",
      journal = {\apjl},
         year = 2021,
        month = jun,
       volume = {914},
       number = {1},
          eid = {L21},
        pages = {L21},
          doi = {10.3847/2041-8213/ac0621},
archivePrefix = {arXiv},
       eprint = {2103.09251},
 primaryClass = {astro-ph.HE},
       adsurl = {https://ui.adsabs.harvard.edu/abs/2021ApJ...914L..21D}
}

@ARTICLE{1964PhRv..136.1224P,
       author = {{Peters}, P.~C.},
        title = "{Gravitational Radiation and the Motion of Two Point Masses}",
      journal = {Physical Review},
         year = 1964,
        month = nov,
       volume = {136},
       number = {4B},
        pages = {1224-1232},
          doi = {10.1103/PhysRev.136.B1224},
       adsurl = {https://ui.adsabs.harvard.edu/abs/1964PhRv..136.1224P}
}

@article{cmocean,
	  author = {Kristen M. Thyng and Chad A. Greene and Robert D. Hetland and Heather M. Zimmerle and Steven F. DiMarco},
	  title = {True Colors of Oceanography: Guidelines for Effective and Accurate Colormap Selection},
	  journal = {Oceanography},
	  year = {2016},
	  month = {September},
	  note = {},
	  url = {https://doi.org/10.5670/oceanog.2016.66},
}

@ARTICLE{1986ApJS...62..519B,
       author = {{Boss}, Alan P.},
        title = "{Protostellar Formation in Rotating Interstellar Clouds. V. Nonisothermal Collapse and Fragmentation}",
      journal = {\apjs},
         year = 1986,
        month = nov,
       volume = {62},
        pages = {519},
          doi = {10.1086/191150},
       adsurl = {https://ui.adsabs.harvard.edu/abs/1986ApJS...62..519B}
}

@ARTICLE{2019MNRAS.489.5822E,
       author = {{El-Badry}, Kareem and {Rix}, Hans-Walter and {Tian}, Haijun and {Duch{\^e}ne}, Gaspard and {Moe}, Maxwell},
        title = "{Discovery of an equal-mass `twin' binary population reaching 1000 + au separations}",
      journal = {\mnras},
         year = 2019,
        month = nov,
       volume = {489},
       number = {4},
        pages = {5822-5857},
          doi = {10.1093/mnras/stz2480},
archivePrefix = {arXiv},
       eprint = {1906.10128},
 primaryClass = {astro-ph.SR},
       adsurl = {https://ui.adsabs.harvard.edu/abs/2019MNRAS.489.5822E}
}

@ARTICLE{2019ApJ...871...84M,
       author = {{Mu{\~n}oz}, Diego J. and {Miranda}, Ryan and {Lai}, Dong},
        title = "{Hydrodynamics of Circumbinary Accretion: Angular Momentum Transfer and Binary Orbital Evolution}",
      journal = {\apj},
         year = 2019,
        month = jan,
       volume = {871},
       number = {1},
          eid = {84},
        pages = {84},
          doi = {10.3847/1538-4357/aaf867},
archivePrefix = {arXiv},
       eprint = {1810.04676},
 primaryClass = {astro-ph.HE},
       adsurl = {https://ui.adsabs.harvard.edu/abs/2019ApJ...871...84M}
}

@ARTICLE{2016ApJ...817...70T,
       author = {{Taylor}, S.~R. and {Huerta}, E.~A. and {Gair}, J.~R. and {McWilliams}, S.~T.},
        title = "{Detecting Eccentric Supermassive Black Hole Binaries with Pulsar Timing Arrays: Resolvable Source Strategies}",
      journal = {\apj},
         year = 2016,
        month = jan,
       volume = {817},
       number = {1},
          eid = {70},
        pages = {70},
          doi = {10.3847/0004-637X/817/1/70},
archivePrefix = {arXiv},
       eprint = {1505.06208},
 primaryClass = {gr-qc},
       adsurl = {https://ui.adsabs.harvard.edu/abs/2016ApJ...817...70T}
}

@ARTICLE{2024ApJ...977..244D,
       author = {{D'Orazio}, Daniel J. and {Duffell}, Paul C. and {Tiede}, Christopher},
        title = "{Fast Methods for Computing Photometric Variability of Eccentric Binaries: Boosting, Lensing, and Variable Accretion}",
      journal = {\apj},
         year = 2024,
        month = dec,
       volume = {977},
       number = {2},
          eid = {244},
        pages = {244},
          doi = {10.3847/1538-4357/ad938b},
archivePrefix = {arXiv},
       eprint = {2403.05629},
 primaryClass = {astro-ph.HE},
       adsurl = {https://ui.adsabs.harvard.edu/abs/2024ApJ...977..244D}
}

@ARTICLE{2022PhRvD.106j3010W,
       author = {{Westernacher-Schneider}, John Ryan and {Zrake}, Jonathan and {MacFadyen}, Andrew and {Haiman}, Zolt{\'a}n},
        title = "{Multiband light curves from eccentric accreting supermassive black hole binaries}",
      journal = {\prd},
         year = 2022,
        month = nov,
       volume = {106},
       number = {10},
          eid = {103010},
        pages = {103010},
          doi = {10.1103/PhysRevD.106.103010},
archivePrefix = {arXiv},
       eprint = {2111.06882},
 primaryClass = {astro-ph.HE},
       adsurl = {https://ui.adsabs.harvard.edu/abs/2022PhRvD.106j3010W}
}

@ARTICLE{2025arXiv250506221M,
       author = {{Mohammed}, Niana N. and {Runnoe}, Jessie C. and {Eracleous}, Michael and {Bogdanovi{\'c}}, Tamara and {Stern}, Daniel and {Simon}, Joseph and {Charisi}, Maria and {Lazio}, T. Joseph W. and {Szekerczes}, Kaitlyn and {Sigurdsson}, Steinn and {Dabbieri}, Collin},
        title = "{Testing the Hypothesis that the Quasar J0950+5128 Harbors a Supermassive Black Hole Binary}",
      journal = {arXiv e-prints},
         year = 2025,
        month = may,
          eid = {arXiv:2505.06221},
        pages = {arXiv:2505.06221},
          doi = {10.48550/arXiv.2505.06221},
archivePrefix = {arXiv},
       eprint = {2505.06221},
 primaryClass = {astro-ph.GA},
       adsurl = {https://ui.adsabs.harvard.edu/abs/2025arXiv250506221M}
}

@ARTICLE{2024PhRvD.109l3544S,
       author = {{Sato-Polito}, Gabriela and {Kamionkowski}, Marc},
        title = "{Exploring the spectrum of stochastic gravitational-wave anisotropies with pulsar timing arrays}",
      journal = {\prd},
         year = 2024,
        month = jun,
       volume = {109},
       number = {12},
          eid = {123544},
        pages = {123544},
          doi = {10.1103/PhysRevD.109.123544},
archivePrefix = {arXiv},
       eprint = {2305.05690},
 primaryClass = {astro-ph.CO},
       adsurl = {https://ui.adsabs.harvard.edu/abs/2024PhRvD.109l3544S}
}

@ARTICLE{2023ApJ...951L...8A,
       author = {{Agazie}, Gabriella and {Anumarlapudi}, Akash and {Archibald}, Anne M. and {Arzoumanian}, Zaven and {Baker}, Paul T. and {B{\'e}csy}, Bence and {Blecha}, Laura and {Brazier}, Adam and {Brook}, Paul R. and {Burke-Spolaor}, Sarah and {Burnette}, Rand and {Case}, Robin and {Charisi}, Maria and {Chatterjee}, Shami and {Chatziioannou}, Katerina and {Cheeseboro}, Belinda D. and {Chen}, Siyuan and {Cohen}, Tyler and {Cordes}, James M. and {Cornish}, Neil J. and {Crawford}, Fronefield and {Cromartie}, H. Thankful and {Crowter}, Kathryn and {Cutler}, Curt J. and {Decesar}, Megan E. and {Degan}, Dallas and {Demorest}, Paul B. and {Deng}, Heling and {Dolch}, Timothy and {Drachler}, Brendan and {Ellis}, Justin A. and {Ferrara}, Elizabeth C. and {Fiore}, William and {Fonseca}, Emmanuel and {Freedman}, Gabriel E. and {Garver-Daniels}, Nate and {Gentile}, Peter A. and {Gersbach}, Kyle A. and {Glaser}, Joseph and {Good}, Deborah C. and {G{\"u}ltekin}, Kayhan and {Hazboun}, Jeffrey S. and {Hourihane}, Sophie and {Islo}, Kristina and {Jennings}, Ross J. and {Johnson}, Aaron D. and {Jones}, Megan L. and {Kaiser}, Andrew R. and {Kaplan}, David L. and {Kelley}, Luke Zoltan and {Kerr}, Matthew and {Key}, Joey S. and {Klein}, Tonia C. and {Laal}, Nima and {Lam}, Michael T. and {Lamb}, William G. and {Lazio}, T. Joseph W. and {Lewandowska}, Natalia and {Littenberg}, Tyson B. and {Liu}, Tingting and {Lommen}, Andrea and {Lorimer}, Duncan R. and {Luo}, Jing and {Lynch}, Ryan S. and {Ma}, Chung-Pei and {Madison}, Dustin R. and {Mattson}, Margaret A. and {McEwen}, Alexander and {McKee}, James W. and {McLaughlin}, Maura A. and {McMann}, Natasha and {Meyers}, Bradley W. and {Meyers}, Patrick M. and {Mingarelli}, Chiara M.~F. and {Mitridate}, Andrea and {Natarajan}, Priyamvada and {Ng}, Cherry and {Nice}, David J. and {Ocker}, Stella Koch and {Olum}, Ken D. and {Pennucci}, Timothy T. and {Perera}, Benetge B.~P. and {Petrov}, Polina and {Pol}, Nihan S. and {Radovan}, Henri A. and {Ransom}, Scott M. and {Ray}, Paul S. and {Romano}, Joseph D. and {Sardesai}, Shashwat C. and {Schmiedekamp}, Ann and {Schmiedekamp}, Carl and {Schmitz}, Kai and {Schult}, Levi and {Shapiro-Albert}, Brent J. and {Siemens}, Xavier and {Simon}, Joseph and {Siwek}, Magdalena S. and {Stairs}, Ingrid H. and {Stinebring}, Daniel R. and {Stovall}, Kevin and {Sun}, Jerry P. and {Susobhanan}, Abhimanyu and {Swiggum}, Joseph K. and {Taylor}, Jacob and {Taylor}, Stephen R. and {Turner}, Jacob E. and {Unal}, Caner and {Vallisneri}, Michele and {van Haasteren}, Rutger and {Vigeland}, Sarah J. and {Wahl}, Haley M. and {Wang}, Qiaohong and {Witt}, Caitlin A. and {Young}, Olivia and {Nanograv Collaboration}},
        title = "{The NANOGrav 15 yr Data Set: Evidence for a Gravitational-wave Background}",
      journal = {\apjl},
         year = 2023,
        month = jul,
       volume = {951},
       number = {1},
          eid = {L8},
        pages = {L8},
          doi = {10.3847/2041-8213/acdac6},
archivePrefix = {arXiv},
       eprint = {2306.16213},
 primaryClass = {astro-ph.HE},
       adsurl = {https://ui.adsabs.harvard.edu/abs/2023ApJ...951L...8A}
}

@ARTICLE{2005ApJ...623..922O,
       author = {{Ochi}, Yasuhiro and {Sugimoto}, Kanako and {Hanawa}, Tomoyuki},
        title = "{Evolution of a Protobinary: Accretion Rates of the Primary and Secondary}",
      journal = {\apj},
         year = 2005,
        month = apr,
       volume = {623},
       number = {2},
        pages = {922-939},
          doi = {10.1086/428601},
       adsurl = {https://ui.adsabs.harvard.edu/abs/2005ApJ...623..922O}
}

@ARTICLE{2010ApJ...708..485H,
       author = {{Hanawa}, Tomoyuki and {Ochi}, Yasuhiro and {Ando}, Koichi},
        title = "{Gas Accretion from a Circumbinary Disk}",
      journal = {\apj},
         year = 2010,
        month = jan,
       volume = {708},
       number = {1},
        pages = {485-497},
          doi = {10.1088/0004-637X/708/1/485},
archivePrefix = {arXiv},
       eprint = {0911.2032},
 primaryClass = {astro-ph.SR},
       adsurl = {https://ui.adsabs.harvard.edu/abs/2010ApJ...708..485H}
}

@ARTICLE{2023ApJ...956L...3A,
       author = {{Agazie}, Gabriella and {Anumarlapudi}, Akash and {Archibald}, Anne M. and {Arzoumanian}, Zaven and {Baker}, Paul T. and {B{\'e}csy}, Bence and {Blecha}, Laura and {Brazier}, Adam and {Brook}, Paul R. and {Burke-Spolaor}, Sarah and {Casey-Clyde}, J. Andrew and {Charisi}, Maria and {Chatterjee}, Shami and {Cohen}, Tyler and {Cordes}, James M. and {Cornish}, Neil J. and {Crawford}, Fronefield and {Cromartie}, H. Thankful and {Crowter}, Kathryn and {DeCesar}, Megan E. and {Demorest}, Paul B. and {Dolch}, Timothy and {Drachler}, Brendan and {Ferrara}, Elizabeth C. and {Fiore}, William and {Fonseca}, Emmanuel and {Freedman}, Gabriel E. and {Gardiner}, Emiko and {Garver-Daniels}, Nate and {Gentile}, Peter A. and {Glaser}, Joseph and {Good}, Deborah C. and {G{\"u}ltekin}, Kayhan and {Hazboun}, Jeffrey S. and {Jennings}, Ross J. and {Johnson}, Aaron D. and {Jones}, Megan L. and {Kaiser}, Andrew R. and {Kaplan}, David L. and {Kelley}, Luke Zoltan and {Kerr}, Matthew and {Key}, Joey S. and {Laal}, Nima and {Lam}, Michael T. and {Lamb}, William G. and {Lazio}, T. Joseph W. and {Lewandowska}, Natalia and {Liu}, Tingting and {Lorimer}, Duncan R. and {Luo}, Jing and {Lynch}, Ryan S. and {Ma}, Chung-Pei and {Madison}, Dustin R. and {McEwen}, Alexander and {McKee}, James W. and {McLaughlin}, Maura A. and {McMann}, Natasha and {Meyers}, Bradley W. and {Mingarelli}, Chiara M.~F. and {Mitridate}, Andrea and {Ng}, Cherry and {Nice}, David J. and {Ocker}, Stella Koch and {Olum}, Ken D. and {Pennucci}, Timothy T. and {Perera}, Benetge B.~P. and {Pol}, Nihan S. and {Radovan}, Henri A. and {Ransom}, Scott M. and {Ray}, Paul S. and {Romano}, Joseph D. and {Sardesai}, Shashwat C. and {Schmiedekamp}, Ann and {Schmiedekamp}, Carl and {Schmitz}, Kai and {Schult}, Levi and {Shapiro-Albert}, Brent J. and {Siemens}, Xavier and {Simon}, Joseph and {Siwek}, Magdalena S. and {Stairs}, Ingrid H. and {Stinebring}, Daniel R. and {Stovall}, Kevin and {Susobhanan}, Abhimanyu and {Swiggum}, Joseph K. and {Taylor}, Stephen R. and {Turner}, Jacob E. and {Unal}, Caner and {Vallisneri}, Michele and {Vigeland}, Sarah J. and {Wahl}, Haley M. and {Witt}, Caitlin A. and {Young}, Olivia},
        title = "{The NANOGrav 15 yr Data Set: Search for Anisotropy in the Gravitational-wave Background}",
      journal = {\apjl},
         year = 2023,
        month = oct,
       volume = {956},
       number = {1},
          eid = {L3},
        pages = {L3},
          doi = {10.3847/2041-8213/acf4fd},
archivePrefix = {arXiv},
       eprint = {2306.16221},
 primaryClass = {astro-ph.HE},
       adsurl = {https://ui.adsabs.harvard.edu/abs/2023ApJ...956L...3A}
}

@ARTICLE{1997MNRAS.285...33B,
       author = {{Bate}, Matthew R. and {Bonnell}, Ian A.},
        title = "{Accretion during binary star formation - II. Gaseous accretion and disc formation}",
      journal = {\mnras},
         year = 1997,
        month = feb,
       volume = {285},
       number = {1},
        pages = {33-48},
          doi = {10.1093/mnras/285.1.33},
       adsurl = {https://ui.adsabs.harvard.edu/abs/1997MNRAS.285...33B}
}

@ARTICLE{2023MNRAS.521.5334P,
       author = {{Park}, Jongwon and {Ricotti}, Massimo and {Sugimura}, Kazuyuki},
        title = "{Population III star formation in an X-ray background: III. Periodic radiative feedback and luminosity induced by elliptical orbits}",
      journal = {\mnras},
         year = 2023,
        month = jun,
       volume = {521},
       number = {4},
        pages = {5334-5353},
          doi = {10.1093/mnras/stad895},
archivePrefix = {arXiv},
       eprint = {2212.04564},
 primaryClass = {astro-ph.GA},
       adsurl = {https://ui.adsabs.harvard.edu/abs/2023MNRAS.521.5334P}
}

@ARTICLE{2023A&A...678A..50E,
       author = {{EPTA Collaboration} and {InPTA Collaboration} and {Antoniadis}, J. and {Arumugam}, P. and {Arumugam}, S. and {Babak}, S. and {Bagchi}, M. and {Bak Nielsen}, A. -S. and {Bassa}, C.~G. and {Bathula}, A. and {Berthereau}, A. and {Bonetti}, M. and {Bortolas}, E. and {Brook}, P.~R. and {Burgay}, M. and {Caballero}, R.~N. and {Chalumeau}, A. and {Champion}, D.~J. and {Chanlaridis}, S. and {Chen}, S. and {Cognard}, I. and {Dandapat}, S. and {Deb}, D. and {Desai}, S. and {Desvignes}, G. and {Dhanda-Batra}, N. and {Dwivedi}, C. and {Falxa}, M. and {Ferdman}, R.~D. and {Franchini}, A. and {Gair}, J.~R. and {Goncharov}, B. and {Gopakumar}, A. and {Graikou}, E. and {Grie{\ss}meier}, J. -M. and {Guillemot}, L. and {Guo}, Y.~J. and {Gupta}, Y. and {Hisano}, S. and {Hu}, H. and {Iraci}, F. and {Izquierdo-Villalba}, D. and {Jang}, J. and {Jawor}, J. and {Janssen}, G.~H. and {Jessner}, A. and {Joshi}, B.~C. and {Kareem}, F. and {Karuppusamy}, R. and {Keane}, E.~F. and {Keith}, M.~J. and {Kharbanda}, D. and {Kikunaga}, T. and {Kolhe}, N. and {Kramer}, M. and {Krishnakumar}, M.~A. and {Lackeos}, K. and {Lee}, K.~J. and {Liu}, K. and {Liu}, Y. and {Lyne}, A.~G. and {McKee}, J.~W. and {Maan}, Y. and {Main}, R.~A. and {Mickaliger}, M.~B. and {Ni{\c{t}}u}, I.~C. and {Nobleson}, K. and {Paladi}, A.~K. and {Parthasarathy}, A. and {Perera}, B.~B.~P. and {Perrodin}, D. and {Petiteau}, A. and {Porayko}, N.~K. and {Possenti}, A. and {Prabu}, T. and {Quelquejay Leclere}, H. and {Rana}, P. and {Samajdar}, A. and {Sanidas}, S.~A. and {Sesana}, A. and {Shaifullah}, G. and {Singha}, J. and {Speri}, L. and {Spiewak}, R. and {Srivastava}, A. and {Stappers}, B.~W. and {Surnis}, M. and {Susarla}, S.~C. and {Susobhanan}, A. and {Takahashi}, K. and {Tarafdar}, P. and {Theureau}, G. and {Tiburzi}, C. and {van der Wateren}, E. and {Vecchio}, A. and {Venkatraman Krishnan}, V. and {Verbiest}, J.~P.~W. and {Wang}, J. and {Wang}, L. and {Wu}, Z.},
        title = "{The second data release from the European Pulsar Timing Array. III. Search for gravitational wave signals}",
      journal = {\aap},
         year = 2023,
        month = oct,
       volume = {678},
          eid = {A50},
        pages = {A50},
          doi = {10.1051/0004-6361/202346844},
archivePrefix = {arXiv},
       eprint = {2306.16214},
 primaryClass = {astro-ph.HE},
       adsurl = {https://ui.adsabs.harvard.edu/abs/2023A&A...678A..50E}
}

@ARTICLE{2017MNRAS.470.1738C,
       author = {{Chen}, Siyuan and {Sesana}, Alberto and {Del Pozzo}, Walter},
        title = "{Efficient computation of the gravitational wave spectrum emitted by eccentric massive black hole binaries in stellar environments}",
      journal = {\mnras},
         year = 2017,
        month = sep,
       volume = {470},
       number = {2},
        pages = {1738-1749},
          doi = {10.1093/mnras/stx1093},
archivePrefix = {arXiv},
       eprint = {1612.00455},
 primaryClass = {astro-ph.CO},
       adsurl = {https://ui.adsabs.harvard.edu/abs/2017MNRAS.470.1738C}
}

@ARTICLE{2021MNRAS.501.2531S,
       author = {{Sayeb}, Mohammad and {Blecha}, Laura and {Kelley}, Luke Zoltan and {Gerosa}, Davide and {Kesden}, Michael and {Thomas}, July},
        title = "{Massive black hole binary inspiral and spin evolution in a cosmological framework}",
      journal = {\mnras},
         year = 2021,
        month = feb,
       volume = {501},
       number = {2},
        pages = {2531-2546},
          doi = {10.1093/mnras/staa3826},
archivePrefix = {arXiv},
       eprint = {2006.06647},
 primaryClass = {astro-ph.GA},
       adsurl = {https://ui.adsabs.harvard.edu/abs/2021MNRAS.501.2531S}
}

@ARTICLE{2024MNRAS.534.3448B,
       author = {{Bourne}, Martin A. and {Fiacconi}, Davide and {Sijacki}, Debora and {Piotrowska}, Joanna M. and {Koudmani}, Sophie},
        title = "{Dynamics and spin alignment in massive, gravito-turbulent circumbinary discs around supermassive black hole binaries}",
      journal = {\mnras},
         year = 2024,
        month = nov,
       volume = {534},
       number = {4},
        pages = {3448-3477},
          doi = {10.1093/mnras/stae2143},
archivePrefix = {arXiv},
       eprint = {2311.17144},
 primaryClass = {astro-ph.HE},
       adsurl = {https://ui.adsabs.harvard.edu/abs/2024MNRAS.534.3448B}
}

@ARTICLE{2025arXiv250816855W,
       author = {{Wang}, Hai-Yang and {Most}, Elias R. and {Hopkins}, Philip F.},
        title = "{$\textit{BMAD}$-Circumbinary Magnetically Arrested Disks around Stellar or Black Hole Binaries: Hot Accretion Flows, Disk Properties, and Angular Momentum Transfer}",
      journal = {arXiv e-prints},
         year = 2025,
        month = aug,
          eid = {arXiv:2508.16855},
        pages = {arXiv:2508.16855},
          doi = {10.48550/arXiv.2508.16855},
archivePrefix = {arXiv},
       eprint = {2508.16855},
 primaryClass = {astro-ph.HE},
       adsurl = {https://ui.adsabs.harvard.edu/abs/2025arXiv250816855W}
}

@ARTICLE{2024ApJ...973L..19M,
       author = {{Most}, Elias R. and {Wang}, Hai-Yang},
        title = "{Magnetically Arrested Circumbinary Accretion Flows}",
      journal = {\apjl},
         year = 2024,
        month = sep,
       volume = {973},
       number = {1},
          eid = {L19},
        pages = {L19},
          doi = {10.3847/2041-8213/ad7713},
archivePrefix = {arXiv},
       eprint = {2408.00757},
 primaryClass = {astro-ph.HE},
       adsurl = {https://ui.adsabs.harvard.edu/abs/2024ApJ...973L..19M}
}

@BOOK{2002apa..book.....F,
       author = {{Frank}, Juhan and {King}, Andrew and {Raine}, Derek J.},
        title = "{Accretion Power in Astrophysics: Third Edition}",
         year = 2002,
       adsurl = {https://ui.adsabs.harvard.edu/abs/2002apa..book.....F},
        publisher = "Cambridge University Press"
}

@ARTICLE{2013ApJ...774..144R,
       author = {{Rafikov}, Roman R.},
        title = "{Structure and Evolution of Circumbinary Disks around Supermassive Black Hole Binaries}",
      journal = {\apj},
         year = 2013,
        month = sep,
       volume = {774},
       number = {2},
          eid = {144},
        pages = {144},
          doi = {10.1088/0004-637X/774/2/144},
archivePrefix = {arXiv},
       eprint = {1205.5017},
 primaryClass = {astro-ph.GA},
       adsurl = {https://ui.adsabs.harvard.edu/abs/2013ApJ...774..144R}
}

@INPROCEEDINGS{1992btsf.work..145L,
       author = {{Lubow}, S.~H. and {Artymowicz}, P.},
        title = "{Eccentricity evolution of a binary embedded in a disk.}",
    booktitle = {Binaries as Tracers of Star Formation},
         year = 1992,
       editor = {{Duquennoy}, Antoine and {Mayor}, Michel},
        month = jan,
        pages = {145-154},
       adsurl = {https://ui.adsabs.harvard.edu/abs/1992btsf.work..145L}
}

@ARTICLE{2024ApJ...975..149M,
       author = {{Marcussen}, Marcus L. and {Albrecht}, Simon H. and {Winn}, Joshua N. and {Su}, Yubo and {Lundkvist}, Mia S. and {Schlaufman}, Kevin C.},
        title = "{The BANANA Project. VII. High Eccentricity Predicts Spin{\textendash}Orbit Misalignment in Binaries}",
      journal = {\apj},
         year = 2024,
        month = nov,
       volume = {975},
       number = {1},
          eid = {149},
        pages = {149},
          doi = {10.3847/1538-4357/ad75fa},
archivePrefix = {arXiv},
       eprint = {2408.03072},
 primaryClass = {astro-ph.SR},
       adsurl = {https://ui.adsabs.harvard.edu/abs/2024ApJ...975..149M}
}

@ARTICLE{2013A&A...551A..50D,
       author = {{Dermine}, T. and {Izzard}, R.~G. and {Jorissen}, A. and {Van Winckel}, H.},
        title = "{Eccentricity-pumping in post-AGB stars with circumbinary discs}",
      journal = {\aap},
         year = 2013,
        month = mar,
       volume = {551},
          eid = {A50},
        pages = {A50},
          doi = {10.1051/0004-6361/201219430},
       adsurl = {https://ui.adsabs.harvard.edu/abs/2013A&A...551A..50D}
}

@INPROCEEDINGS{2000prpl.conf..731L,
       author = {{Lubow}, S.~H. and {Artymowicz}, P.},
        title = "{Interactions of Young Binaries with Disks}",
    booktitle = {Protostars and Planets IV},
         year = 2000,
       editor = {{Mannings}, V. and {Boss}, A.~P. and {Russell}, S.~S.},
        month = may,
        pages = {731},
          doi = {10.48550/arXiv.2111.07411},
archivePrefix = {arXiv},
       eprint = {2111.07411},
 primaryClass = {astro-ph.SR},
       adsurl = {https://ui.adsabs.harvard.edu/abs/2000prpl.conf..731L}
}

@ARTICLE{2008ApJ...672...83M,
       author = {{MacFadyen}, Andrew I. and {Milosavljevi{\'c}}, Milo{\v{s}}},
        title = "{An Eccentric Circumbinary Accretion Disk and the Detection of Binary Massive Black Holes}",
      journal = {\apj},
         year = 2008,
        month = jan,
       volume = {672},
       number = {1},
        pages = {83-93},
          doi = {10.1086/523869},
archivePrefix = {arXiv},
       eprint = {astro-ph/0607467},
 primaryClass = {astro-ph},
       adsurl = {https://ui.adsabs.harvard.edu/abs/2008ApJ...672...83M}
}

@ARTICLE{1998A&A...332..877J,
       author = {{Jorissen}, A. and {Van Eck}, S. and {Mayor}, M. and {Udry}, S.},
        title = "{Insights into the formation of barium and Tc-poor S stars from an extended sample of orbital elements}",
      journal = {\aap},
         year = 1998,
        month = apr,
       volume = {332},
        pages = {877-903},
          doi = {10.48550/arXiv.astro-ph/9801272},
archivePrefix = {arXiv},
       eprint = {astro-ph/9801272},
 primaryClass = {astro-ph},
       adsurl = {https://ui.adsabs.harvard.edu/abs/1998A&A...332..877J}
}

@ARTICLE{2018A&A...620A..85O,
       author = {{Oomen}, Glenn-Michael and {Van Winckel}, Hans and {Pols}, Onno and {Nelemans}, Gijs and {Escorza}, Ana and {Manick}, Rajeev and {Kamath}, Devika and {Waelkens}, Christoffel},
        title = "{Orbital properties of binary post-AGB stars}",
      journal = {\aap},
         year = 2018,
        month = dec,
       volume = {620},
          eid = {A85},
        pages = {A85},
          doi = {10.1051/0004-6361/201833816},
archivePrefix = {arXiv},
       eprint = {1810.01842},
 primaryClass = {astro-ph.SR},
       adsurl = {https://ui.adsabs.harvard.edu/abs/2018A&A...620A..85O}
}

@ARTICLE{1980Natur.287..307B,
       author = {{Begelman}, M.~C. and {Blandford}, R.~D. and {Rees}, M.~J.},
        title = "{Massive black hole binaries in active galactic nuclei}",
      journal = {\nat},
         year = 1980,
        month = sep,
       volume = {287},
       number = {5780},
        pages = {307-309},
          doi = {10.1038/287307a0},
       adsurl = {https://ui.adsabs.harvard.edu/abs/1980Natur.287..307B}
}

@ARTICLE{2015MNRAS.447.2907Y,
       author = {{Young}, M.~D. and {Baird}, J.~T. and {Clarke}, C.~J.},
        title = "{The evolution of the mass ratio of accreting binaries: the role of gas temperature}",
      journal = {\mnras},
         year = 2015,
        month = mar,
       volume = {447},
       number = {3},
        pages = {2907-2914},
          doi = {10.1093/mnras/stu2656},
archivePrefix = {arXiv},
       eprint = {1412.3963},
 primaryClass = {astro-ph.SR},
       adsurl = {https://ui.adsabs.harvard.edu/abs/2015MNRAS.447.2907Y}
}

@ARTICLE{2025ApJ...984..144T,
       author = {{Tiede}, Christopher and {Zrake}, Jonathan and {MacFadyen}, Andrew and {Haiman}, Zolt{\'a}n},
        title = "{Suppressed Accretion onto Massive Black Hole Binaries Surrounded by Thin Disks}",
      journal = {\apj},
         year = 2025,
        month = may,
       volume = {984},
       number = {2},
          eid = {144},
        pages = {144},
          doi = {10.3847/1538-4357/adc727},
archivePrefix = {arXiv},
       eprint = {2410.03830},
 primaryClass = {astro-ph.GA},
       adsurl = {https://ui.adsabs.harvard.edu/abs/2025ApJ...984..144T}
}

@ARTICLE{2019Sci...366...90A,
       author = {{Alves}, F.~O. and {Caselli}, P. and {Girart}, J.~M. and {Segura-Cox}, D. and {Franco}, G.~A.~P. and {Schmiedeke}, A. and {Zhao}, B.},
        title = "{Gas flow and accretion via spiral streamers and circumstellar disks in a young binary protostar}",
      journal = {Science},
         year = 2019,
        month = oct,
       volume = {366},
       number = {6461},
        pages = {90-93},
          doi = {10.1126/science.aaw3491},
archivePrefix = {arXiv},
       eprint = {1910.01141},
 primaryClass = {astro-ph.SR},
       adsurl = {https://ui.adsabs.harvard.edu/abs/2019Sci...366...90A}
}

@ARTICLE{2008ApJ...680..169S,
       author = {{Shen}, Yue and {Greene}, Jenny E. and {Strauss}, Michael A. and {Richards}, Gordon T. and {Schneider}, Donald P.},
        title = "{Biases in Virial Black Hole Masses: An SDSS Perspective}",
      journal = {\apj},
         year = 2008,
        month = jun,
       volume = {680},
       number = {1},
        pages = {169-190},
          doi = {10.1086/587475},
archivePrefix = {arXiv},
       eprint = {0709.3098},
 primaryClass = {astro-ph},
       adsurl = {https://ui.adsabs.harvard.edu/abs/2008ApJ...680..169S}
}

@ARTICLE{2024ApJ...969..132V,
       author = {{Vijaykumar}, Aditya and {Hanselman}, Alexandra G. and {Zevin}, Michael},
        title = "{Consistent Eccentricities for Gravitational-wave Astronomy: Resolving Discrepancies between Astrophysical Simulations and Waveform Models}",
      journal = {\apj},
         year = 2024,
        month = jul,
       volume = {969},
       number = {2},
          eid = {132},
        pages = {132},
          doi = {10.3847/1538-4357/ad4455},
archivePrefix = {arXiv},
       eprint = {2402.07892},
 primaryClass = {astro-ph.HE},
       adsurl = {https://ui.adsabs.harvard.edu/abs/2024ApJ...969..132V}
}

@ARTICLE{2025ApJ...982L..34W,
       author = {{Wu}, Yanqin and {Hadden}, Sam and {Dewberry}, Janosz and {El-Badry}, Kareem and {Matzner}, Christopher D.},
        title = "{Eccentricities of Close Stellar Binaries}",
      journal = {\apjl},
         year = 2025,
        month = mar,
       volume = {982},
       number = {1},
          eid = {L34},
        pages = {L34},
          doi = {10.3847/2041-8213/adb751},
archivePrefix = {arXiv},
       eprint = {2411.09905},
 primaryClass = {astro-ph.SR},
       adsurl = {https://ui.adsabs.harvard.edu/abs/2025ApJ...982L..34W}
}

@ARTICLE{2023LRR....26....2A,
       author = {{Amaro-Seoane}, Pau and {Andrews}, Jeff and {Arca Sedda}, Manuel and {Askar}, Abbas and {Baghi}, Quentin and {Balasov}, Razvan and {Bartos}, Imre and {Bavera}, Simone S. and {Bellovary}, Jillian and {Berry}, Christopher P.~L. and {Berti}, Emanuele and {Bianchi}, Stefano and {Blecha}, Laura and {Blondin}, St{\'e}phane and {Bogdanovi{\'c}}, Tamara and {Boissier}, Samuel and {Bonetti}, Matteo and {Bonoli}, Silvia and {Bortolas}, Elisa and {Breivik}, Katelyn and {Capelo}, Pedro R. and {Caramete}, Laurentiu and {Cattorini}, Federico and {Charisi}, Maria and {Chaty}, Sylvain and {Chen}, Xian and {Chru{\'s}li{\'n}ska}, Martyna and {Chua}, Alvin J.~K. and {Church}, Ross and {Colpi}, Monica and {D'Orazio}, Daniel and {Danielski}, Camilla and {Davies}, Melvyn B. and {Dayal}, Pratika and {De Rosa}, Alessandra and {Derdzinski}, Andrea and {Destounis}, Kyriakos and {Dotti}, Massimo and {Du{\c{t}}an}, Ioana and {Dvorkin}, Irina and {Fabj}, Gaia and {Foglizzo}, Thierry and {Ford}, Saavik and {Fouvry}, Jean-Baptiste and {Franchini}, Alessia and {Fragos}, Tassos and {Fryer}, Chris and {Gaspari}, Massimo and {Gerosa}, Davide and {Graziani}, Luca and {Groot}, Paul and {Habouzit}, Melanie and {Haggard}, Daryl and {Haiman}, Zoltan and {Han}, Wen-Biao and {Istrate}, Alina and {Johansson}, Peter H. and {Khan}, Fazeel Mahmood and {Kimpson}, Tomas and {Kokkotas}, Kostas and {Kong}, Albert and {Korol}, Valeriya and {Kremer}, Kyle and {Kupfer}, Thomas and {Lamberts}, Astrid and {Larson}, Shane and {Lau}, Mike and {Liu}, Dongliang and {Lloyd-Ronning}, Nicole and {Lodato}, Giuseppe and {Lupi}, Alessandro and {Ma}, Chung-Pei and {Maccarone}, Tomas and {Mandel}, Ilya and {Mangiagli}, Alberto and {Mapelli}, Michela and {Mathis}, St{\'e}phane and {Mayer}, Lucio and {McGee}, Sean and {McKernan}, Berry and {Miller}, M. Coleman and {Mota}, David F. and {Mumpower}, Matthew and {Nasim}, Syeda S. and {Nelemans}, Gijs and {Noble}, Scott and {Pacucci}, Fabio and {Panessa}, Francesca and {Paschalidis}, Vasileios and {Pfister}, Hugo and {Porquet}, Delphine and {Quenby}, John and {Ricarte}, Angelo and {R{\"o}pke}, Friedrich K. and {Regan}, John and {Rosswog}, Stephan and {Ruiter}, Ashley and {Ruiz}, Milton and {Runnoe}, Jessie and {Schneider}, Raffaella and {Schnittman}, Jeremy and {Secunda}, Amy and {Sesana}, Alberto and {Seto}, Naoki and {Shao}, Lijing and {Shapiro}, Stuart and {Sopuerta}, Carlos and {Stone}, Nicholas C. and {Suvorov}, Arthur and {Tamanini}, Nicola and {Tamfal}, Tomas and {Tauris}, Thomas and {Temmink}, Karel and {Tomsick}, John and {Toonen}, Silvia and {Torres-Orjuela}, Alejandro and {Toscani}, Martina and {Tsokaros}, Antonios and {Unal}, Caner and {V{\'a}zquez-Aceves}, Ver{\'o}nica and {Valiante}, Rosa and {van Putten}, Maurice and {van Roestel}, Jan and {Vignali}, Christian and {Volonteri}, Marta and {Wu}, Kinwah and {Younsi}, Ziri and {Yu}, Shenghua and {Zane}, Silvia and {Zwick}, Lorenz and {Antonini}, Fabio and {Baibhav}, Vishal and {Barausse}, Enrico and {Bonilla Rivera}, Alexander and {Branchesi}, Marica and {Branduardi-Raymont}, Graziella and {Burdge}, Kevin and {Chakraborty}, Srija and {Cuadra}, Jorge and {Dage}, Kristen and {Davis}, Benjamin and {de Mink}, Selma E. and {Decarli}, Roberto and {Doneva}, Daniela and {Escoffier}, Stephanie and {Gandhi}, Poshak and {Haardt}, Francesco and {Lousto}, Carlos O. and {Nissanke}, Samaya and {Nordhaus}, Jason and {O'Shaughnessy}, Richard and {Portegies Zwart}, Simon and {Pound}, Adam and {Schussler}, Fabian and {Sergijenko}, Olga and {Spallicci}, Alessandro and {Vernieri}, Daniele and {Vigna-G{\'o}mez}, Alejandro},
        title = "{Astrophysics with the Laser Interferometer Space Antenna}",
      journal = {Living Reviews in Relativity},
         year = 2023,
        month = dec,
       volume = {26},
       number = {1},
          eid = {2},
        pages = {2},
          doi = {10.1007/s41114-022-00041-y},
archivePrefix = {arXiv},
       eprint = {2203.06016},
 primaryClass = {gr-qc},
       adsurl = {https://ui.adsabs.harvard.edu/abs/2023LRR....26....2A}
}

@INPROCEEDINGS{2023ASPC..534..275O,
       author = {{Offner}, S.~S.~R. and {Moe}, M. and {Kratter}, K.~M. and {Sadavoy}, S.~I. and {Jensen}, E.~L.~N. and {Tobin}, J.~J.},
        title = "{The Origin and Evolution of Multiple Star Systems}",
    booktitle = {Protostars and Planets VII},
         year = 2023,
       editor = {{Inutsuka}, S. and {Aikawa}, Y. and {Muto}, T. and {Tomida}, K. and {Tamura}, M.},
       series = {Astronomical Society of the Pacific Conference Series},
       volume = {534},
        month = jul,
        pages = {275},
          doi = {10.48550/arXiv.2203.10066},
archivePrefix = {arXiv},
       eprint = {2203.10066},
 primaryClass = {astro-ph.SR},
       adsurl = {https://ui.adsabs.harvard.edu/abs/2023ASPC..534..275O}
}

@ARTICLE{2016Natur.538..483T,
       author = {{Tobin}, John J. and {Kratter}, Kaitlin M. and {Persson}, Magnus V. and {Looney}, Leslie W. and {Dunham}, Michael M. and {Segura-Cox}, Dominique and {Li}, Zhi-Yun and {Chandler}, Claire J. and {Sadavoy}, Sarah I. and {Harris}, Robert J. and {Melis}, Carl and {P{\'e}rez}, Laura M.},
        title = "{A triple protostar system formed via fragmentation of a gravitationally unstable disk}",
      journal = {\nat},
         year = 2016,
        month = oct,
       volume = {538},
       number = {7626},
        pages = {483-486},
          doi = {10.1038/nature20094},
archivePrefix = {arXiv},
       eprint = {1610.08524},
 primaryClass = {astro-ph.SR},
       adsurl = {https://ui.adsabs.harvard.edu/abs/2016Natur.538..483T}
}

@ARTICLE{2009A&A...505.1221V,
       author = {{van Winckel}, H. and {Lloyd Evans}, T. and {Briquet}, M. and {De Cat}, P. and {Degroote}, P. and {De Meester}, W. and {De Ridder}, J. and {Deroo}, P. and {Desmet}, M. and {Drummond}, R. and {Eyer}, L. and {Groenewegen}, M.~A.~T. and {Kolenberg}, K. and {Kilkenny}, D. and {Ladjal}, D. and {Lefever}, K. and {Maas}, T. and {Marang}, F. and {Martinez}, P. and {{\O}stensen}, R.~H. and {Raskin}, G. and {Reyniers}, M. and {Royer}, P. and {Saesen}, S. and {Uytterhoeven}, K. and {Vanautgaerden}, J. and {Vandenbussche}, B. and {van Wyk}, F. and {Vu{\v{c}}kovi{\'c}}, M. and {Waelkens}, C. and {Zima}, W.},
        title = "{Post-AGB stars with hot circumstellar dust: binarity of the low-amplitude pulsators}",
      journal = {\aap},
         year = 2009,
        month = oct,
       volume = {505},
       number = {3},
        pages = {1221-1232},
          doi = {10.1051/0004-6361/200912332},
archivePrefix = {arXiv},
       eprint = {0906.4482},
 primaryClass = {astro-ph.SR},
       adsurl = {https://ui.adsabs.harvard.edu/abs/2009A&A...505.1221V}
}

@ARTICLE{2021RNAAS...5..275H,
       author = {{Hamers}, Adrian S.},
        title = "{An Improved Numerical Fit to the Peak Harmonic Gravitational Wave Frequency Emitted by an Eccentric Binary}",
      journal = {Research Notes of the American Astronomical Society},
         year = 2021,
        month = nov,
       volume = {5},
       number = {11},
          eid = {275},
        pages = {275},
          doi = {10.3847/2515-5172/ac3d98},
archivePrefix = {arXiv},
       eprint = {2111.08033},
 primaryClass = {gr-qc},
       adsurl = {https://ui.adsabs.harvard.edu/abs/2021RNAAS...5..275H}
}

@ARTICLE{2020A&A...639A..62K,
       author = {{Keppler}, M. and {Penzlin}, A. and {Benisty}, M. and {van Boekel}, R. and {Henning}, T. and {van Holstein}, R.~G. and {Kley}, W. and {Garufi}, A. and {Ginski}, C. and {Brandner}, W. and {Bertrang}, G.~H.-M. and {Boccaletti}, A. and {de Boer}, J. and {Bonavita}, M. and {Brown Sevilla}, S. and {Chauvin}, G. and {Dominik}, C. and {Janson}, M. and {Langlois}, M. and {Lodato}, G. and {Maire}, A.-L. and {M{\'e}nard}, F. and {Pantin}, E. and {Pinte}, C. and {Stolker}, T. and {Szul{\'a}gyi}, J. and {Thebault}, P. and {Villenave}, M. and {Zurlo}, A. and {Rabou}, P. and {Feautrier}, P. and {Feldt}, M. and {Madec}, F. and {Wildi}, F.},
        title = "{Gap, shadows, spirals, and streamers: SPHERE observations of binary-disk interactions in GG Tauri A}",
      journal = {\aap},
         year = 2020,
        month = jul,
       volume = {639},
          eid = {A62},
        pages = {A62},
          doi = {10.1051/0004-6361/202038032},
archivePrefix = {arXiv},
       eprint = {2005.09037},
 primaryClass = {astro-ph.SR},
       adsurl = {https://ui.adsabs.harvard.edu/abs/2020A&A...639A..62K}
}

@ARTICLE{2021NatAs...5..881G,
       author = {{Gong}, Yungui and {Luo}, Jun and {Wang}, Bin},
        title = "{Concepts and status of Chinese space gravitational wave detection projects}",
      journal = {Nature Astronomy},
         year = 2021,
        month = sep,
       volume = {5},
        pages = {881-889},
          doi = {10.1038/s41550-021-01480-3},
archivePrefix = {arXiv},
       eprint = {2109.07442},
 primaryClass = {astro-ph.IM},
       adsurl = {https://ui.adsabs.harvard.edu/abs/2021NatAs...5..881G}
}

@ARTICLE{2006A&A...448..641D,
       author = {{de Ruyter}, S. and {van Winckel}, H. and {Maas}, T. and {Lloyd Evans}, T. and {Waters}, L.~B.~F.~M. and {Dejonghe}, H.},
        title = "{Keplerian discs around post-AGB stars: a common phenomenon?}",
      journal = {\aap},
         year = 2006,
        month = mar,
       volume = {448},
       number = {2},
        pages = {641-653},
          doi = {10.1051/0004-6361:20054062},
archivePrefix = {arXiv},
       eprint = {astro-ph/0601578},
 primaryClass = {astro-ph},
       adsurl = {https://ui.adsabs.harvard.edu/abs/2006A&A...448..641D}
}

@book{kant1755allgemeine,
  title={Allgemeine Naturgeschichte und Theorie des Himmels, nach Newtonischen Grunds{\"a}tzen abgehandelt},
  author={Kant, I.},
  url={},
  year={1755},
  publisher={ }
}

@ARTICLE{2022MNRAS.513.6158D,
       author = {{Dittmann}, Alexander J. and {Ryan}, Geoffrey},
        title = "{A survey of disc thickness and viscosity in circumbinary accretion: Binary evolution, variability, and disc morphology}",
      journal = {\mnras},
         year = 2022,
        month = jul,
       volume = {513},
       number = {4},
        pages = {6158-6176},
          doi = {10.1093/mnras/stac935},
archivePrefix = {arXiv},
       eprint = {2201.07816},
 primaryClass = {astro-ph.HE},
       adsurl = {https://ui.adsabs.harvard.edu/abs/2022MNRAS.513.6158D}
}

@ARTICLE{2005ApJ...620L..79S,
       author = {{Springel}, Volker and {Di Matteo}, Tiziana and {Hernquist}, Lars},
        title = "{Black Holes in Galaxy Mergers: The Formation of Red Elliptical Galaxies}",
      journal = {\apjl},
         year = 2005,
        month = feb,
       volume = {620},
       number = {2},
        pages = {L79-L82},
          doi = {10.1086/428772},
archivePrefix = {arXiv},
       eprint = {astro-ph/0409436},
 primaryClass = {astro-ph},
       adsurl = {https://ui.adsabs.harvard.edu/abs/2005ApJ...620L..79S}
}

@ARTICLE{1991ApJ...370L..65B,
       author = {{Barnes}, Joshua E. and {Hernquist}, Lars E.},
        title = "{Fueling Starburst Galaxies with Gas-rich Mergers}",
      journal = {\apjl},
         year = 1991,
        month = apr,
       volume = {370},
        pages = {L65},
          doi = {10.1086/185978},
       adsurl = {https://ui.adsabs.harvard.edu/abs/1991ApJ...370L..65B}
}

@ARTICLE{2023ARA&A..61..517L,
       author = {{Lai}, Dong and {Mu{\~n}oz}, Diego J.},
        title = "{Circumbinary Accretion: From Binary Stars to Massive Binary Black Holes}",
      journal = {\araa},
         year = 2023,
        month = aug,
       volume = {61},
        pages = {517-560},
          doi = {10.1146/annurev-astro-052622-022933},
archivePrefix = {arXiv},
       eprint = {2211.00028},
 primaryClass = {astro-ph.HE},
       adsurl = {https://ui.adsabs.harvard.edu/abs/2023ARA&A..61..517L}
}

@ARTICLE{2013CQGra..30x4008M,
       author = {{Mayer}, Lucio},
        title = "{Massive black hole binaries in gas-rich galaxy mergers; multiple regimes of orbital decay and interplay with gas inflows}",
      journal = {Classical and Quantum Gravity},
         year = 2013,
        month = dec,
       volume = {30},
       number = {24},
          eid = {244008},
        pages = {244008},
          doi = {10.1088/0264-9381/30/24/244008},
archivePrefix = {arXiv},
       eprint = {1308.0431},
 primaryClass = {astro-ph.CO},
       adsurl = {https://ui.adsabs.harvard.edu/abs/2013CQGra..30x4008M}
}

@ARTICLE{2024ApJ...970..156D,
       author = {{Duffell}, Paul C. and {Dittmann}, Alexander J. and {D'Orazio}, Daniel J. and {Franchini}, Alessia and {Kratter}, Kaitlin M. and {Penzlin}, Anna B.~T. and {Ragusa}, Enrico and {Siwek}, Magdalena and {Tiede}, Christopher and {Wang}, Haiyang and {Zrake}, Jonathan and {Dempsey}, Adam M. and {Haiman}, Zoltan and {Lupi}, Alessandro and {Pirog}, Michal and {Ryan}, Geoffrey},
        title = "{The Santa Barbara Binary‑disk Code Comparison}",
      journal = {\apj},
         year = 2024,
        month = aug,
       volume = {970},
       number = {2},
          eid = {156},
        pages = {156},
          doi = {10.3847/1538-4357/ad5a7e},
archivePrefix = {arXiv},
       eprint = {2402.13039},
 primaryClass = {astro-ph.SR},
       adsurl = {https://ui.adsabs.harvard.edu/abs/2024ApJ...970..156D}
}

@ARTICLE{2021ApJ...921...71D,
       author = {{Dittmann}, Alexander J. and {Ryan}, Geoffrey},
        title = "{Preventing Anomalous Torques in Circumbinary Accretion Simulations}",
      journal = {\apj},
         year = 2021,
        month = nov,
       volume = {921},
       number = {1},
          eid = {71},
        pages = {71},
          doi = {10.3847/1538-4357/ac1bbd},
archivePrefix = {arXiv},
       eprint = {2102.05684},
 primaryClass = {astro-ph.HE},
       adsurl = {https://ui.adsabs.harvard.edu/abs/2021ApJ...921...71D}
}

@ARTICLE{2024ApJ...967...12D,
       author = {{Dittmann}, Alexander J. and {Ryan}, Geoffrey},
        title = "{The Evolution of Accreting Binaries: From Brown Dwarfs to Supermassive Black Holes}",
      journal = {\apj},
         year = 2024,
        month = may,
       volume = {967},
       number = {1},
          eid = {12},
        pages = {12},
          doi = {10.3847/1538-4357/ad2f1e},
archivePrefix = {arXiv},
       eprint = {2310.07758},
 primaryClass = {astro-ph.GA},
       adsurl = {https://ui.adsabs.harvard.edu/abs/2024ApJ...967...12D}
}

@ARTICLE{1979JCoPh..32..101V,
       author = {{van Leer}, B.},
        title = "{Towards the Ultimate Conservative Difference Scheme. V. A Second-Order Sequel to Godunov's Method}",
      journal = {Journal of Computational Physics},
         year = 1979,
        month = jul,
       volume = {32},
       number = {1},
        pages = {101-136},
          doi = {10.1016/0021-9991(79)90145-1},
       adsurl = {https://ui.adsabs.harvard.edu/abs/1979JCoPh..32..101V}
}

@ARTICLE{2000JCoPh.160..241K,
       author = {{Kurganov}, Alexander and {Tadmor}, Eitan},
        title = "{New High-Resolution Central Schemes for Nonlinear Conservation Laws and Convection-Diffusion Equations}",
      journal = {Journal of Computational Physics},
         year = 2000,
        month = may,
       volume = {160},
       number = {1},
        pages = {241-282},
          doi = {10.1006/jcph.2000.6459},
       adsurl = {https://ui.adsabs.harvard.edu/abs/2000JCoPh.160..241K}
}

@ARTICLE{1994ShWav...4...25T,
       author = {{Toro}, E.~F. and {Spruce}, M. and {Speares}, W.},
        title = "{Restoration of the contact surface in the HLL-Riemann solver}",
      journal = {Shock Waves},
         year = 1994,
        month = jul,
       volume = {4},
       number = {1},
        pages = {25-34},
          doi = {10.1007/BF01414629},
       adsurl = {https://ui.adsabs.harvard.edu/abs/1994ShWav...4...25T}
}

@ARTICLE{1998MaCom..67...73G,
       author = {{Gottlieb}, S. and {Shu}, C.~W.},
        title = "{Total variation diminishing Runge-Kutta schemes}",
      journal = {Mathematics of Computation},
         year = 1998,
        month = jan,
       volume = {67},
       number = {221},
        pages = {73-85},
       adsurl = {https://ui.adsabs.harvard.edu/abs/1998MaCom..67...73G}
}

@ARTICLE{2021ApJ...922..175N,
       author = {{Noble}, Scott C. and {Krolik}, Julian H. and {Campanelli}, Manuela and {Zlochower}, Yosef and {Mundim}, Bruno C. and {Nakano}, Hiroyuki and {Zilh{\~a}o}, Miguel},
        title = "{Mass-ratio and Magnetic Flux Dependence of Modulated Accretion from Circumbinary Disks}",
      journal = {\apj},
         year = 2021,
        month = dec,
       volume = {922},
       number = {2},
          eid = {175},
        pages = {175},
          doi = {10.3847/1538-4357/ac2229},
archivePrefix = {arXiv},
       eprint = {2103.12100},
 primaryClass = {astro-ph.HE},
       adsurl = {https://ui.adsabs.harvard.edu/abs/2021ApJ...922..175N}
}

@ARTICLE{2025ApJ...986..158T,
       author = {{Tiwari}, Vishal and {Chan}, Chi-Ho and {Bogdanovi{\'c}}, Tamara and {Jiang}, Yan-Fei and {Davis}, Shane W. and {Ferrel}, Simon},
        title = "{Radiation Magnetohydrodynamic Simulation of Sub-Eddington Circumbinary Disk around an Equal-mass Massive Black Hole Binary}",
      journal = {\apj},
         year = 2025,
        month = jun,
       volume = {986},
       number = {2},
          eid = {158},
        pages = {158},
          doi = {10.3847/1538-4357/add408},
archivePrefix = {arXiv},
       eprint = {2502.18584},
 primaryClass = {astro-ph.HE},
       adsurl = {https://ui.adsabs.harvard.edu/abs/2025ApJ...986..158T}
}

@ARTICLE{2025MNRAS.537.3620Z,
       author = {{Zrake}, Jonathan and {Clyburn}, Madeline and {Feyan}, Samuel},
        title = "{Changing-look inspirals: trends and switches in AGN disc emission as signposts for merging black hole binaries}",
      journal = {\mnras},
         year = 2025,
        month = mar,
       volume = {537},
       number = {4},
        pages = {3620-3631},
          doi = {10.1093/mnras/staf171},
archivePrefix = {arXiv},
       eprint = {2410.04961},
 primaryClass = {astro-ph.HE},
       adsurl = {https://ui.adsabs.harvard.edu/abs/2025MNRAS.537.3620Z}
}

@ARTICLE{2023MNRAS.526.5441K,
       author = {{Krauth}, Luke Major and {Davelaar}, Jordy and {Haiman}, Zolt{\'a}n and {Westernacher-Schneider}, John Ryan and {Zrake}, Jonathan and {MacFadyen}, Andrew},
        title = "{Disappearing thermal X-ray emission as a tell-tale signature of merging massive black hole binaries}",
      journal = {\mnras},
         year = 2023,
        month = dec,
       volume = {526},
       number = {4},
        pages = {5441-5454},
          doi = {10.1093/mnras/stad3095},
archivePrefix = {arXiv},
       eprint = {2304.02575},
 primaryClass = {astro-ph.HE},
       adsurl = {https://ui.adsabs.harvard.edu/abs/2023MNRAS.526.5441K}
}

@ARTICLE{2023arXiv231016896D,
       author = {{D'Orazio}, Daniel J. and {Charisi}, Maria},
        title = "{Observational Signatures of Supermassive Black Hole Binaries}",
      journal = {arXiv e-prints},
         year = 2023,
        month = oct,
          eid = {arXiv:2310.16896},
        pages = {arXiv:2310.16896},
          doi = {10.48550/arXiv.2310.16896},
archivePrefix = {arXiv},
       eprint = {2310.16896},
 primaryClass = {astro-ph.HE},
       adsurl = {https://ui.adsabs.harvard.edu/abs/2023arXiv231016896D}
}

@ARTICLE{2024MNRAS.532..295F,
       author = {{Fastidio}, F. and {Gualandris}, A. and {Sesana}, A. and {Bortolas}, E. and {Dehnen}, W.},
        title = "{Eccentricity evolution of PTA sources from cosmological initial conditions}",
      journal = {\mnras},
         year = 2024,
        month = jul,
       volume = {532},
       number = {1},
        pages = {295-304},
          doi = {10.1093/mnras/stae1411},
archivePrefix = {arXiv},
       eprint = {2406.02710},
 primaryClass = {astro-ph.GA},
       adsurl = {https://ui.adsabs.harvard.edu/abs/2024MNRAS.532..295F}
}

@ARTICLE{1995A&A...296..709V,
       author = {{Verbunt}, F. and {Phinney}, E.~S.},
        title = "{Tidal circularization and the eccentricity of binaries containing giant stars.}",
      journal = {\aap},
         year = 1995,
        month = apr,
       volume = {296},
        pages = {709},
       adsurl = {https://ui.adsabs.harvard.edu/abs/1995A&A...296..709V}
}

@ARTICLE{2018ApJ...867....5P,
       author = {{Price-Whelan}, Adrian M. and {Goodman}, Jeremy},
        title = "{Binary Companions of Evolved Stars in APOGEE DR14: Orbital Circularization}",
      journal = {\apj},
         year = 2018,
        month = nov,
       volume = {867},
       number = {1},
          eid = {5},
        pages = {5},
          doi = {10.3847/1538-4357/aae264},
archivePrefix = {arXiv},
       eprint = {1804.06841},
 primaryClass = {astro-ph.SR},
       adsurl = {https://ui.adsabs.harvard.edu/abs/2018ApJ...867....5P}
}

@ARTICLE{2012ApJ...749..118S,
       author = {{Shi}, Ji-Ming and {Krolik}, Julian H. and {Lubow}, Stephen H. and {Hawley}, John F.},
        title = "{Three-dimensional Magnetohydrodynamic Simulations of Circumbinary Accretion Disks: Disk Structures and Angular Momentum Transport}",
      journal = {\apj},
         year = 2012,
        month = apr,
       volume = {749},
       number = {2},
          eid = {118},
        pages = {118},
          doi = {10.1088/0004-637X/749/2/118},
archivePrefix = {arXiv},
       eprint = {1110.4866},
 primaryClass = {astro-ph.HE},
       adsurl = {https://ui.adsabs.harvard.edu/abs/2012ApJ...749..118S}
}

@ARTICLE{2025arXiv250206389E,
       author = {{Ennoggi}, Lorenzo and {Campanelli}, Manuela and {Zlochower}, Yosef and {Noble}, Scott C. and {Krolik}, Julian and {Cattorini}, Federico and {Kalinani}, Jay V. and {Mewes}, Vassilios and {Chabanov}, Michail and {Ji}, Liwei and {de Simone}, Maria Chiara},
        title = "{Relativistic gas accretion onto supermassive black Hole binaries from inspiral through merger}",
      journal = {arXiv e-prints},
         year = 2025,
        month = feb,
          eid = {arXiv:2502.06389},
        pages = {arXiv:2502.06389},
          doi = {10.48550/arXiv.2502.06389},
archivePrefix = {arXiv},
       eprint = {2502.06389},
 primaryClass = {astro-ph.HE},
       adsurl = {https://ui.adsabs.harvard.edu/abs/2025arXiv250206389E}
}

@ARTICLE{2023ApJ...949L..30D,
       author = {{Dittmann}, Alexander J. and {Ryan}, Geoffrey and {Miller}, M. Coleman},
        title = "{The Decoupling of Binaries from Their Circumbinary Disks}",
      journal = {\apjl},
         year = 2023,
        month = jun,
       volume = {949},
       number = {2},
          eid = {L30},
        pages = {L30},
          doi = {10.3847/2041-8213/acd183},
archivePrefix = {arXiv},
       eprint = {2303.16204},
 primaryClass = {astro-ph.HE},
       adsurl = {https://ui.adsabs.harvard.edu/abs/2023ApJ...949L..30D}
}

@ARTICLE{2002ApJ...567L...9A,
       author = {{Armitage}, Philip J. and {Natarajan}, Priyamvada},
        title = "{Accretion during the Merger of Supermassive Black Holes}",
      journal = {\apjl},
         year = 2002,
        month = mar,
       volume = {567},
       number = {1},
        pages = {L9-L12},
          doi = {10.1086/339770},
archivePrefix = {arXiv},
       eprint = {astro-ph/0201318},
 primaryClass = {astro-ph},
       adsurl = {https://ui.adsabs.harvard.edu/abs/2002ApJ...567L...9A}
}

\appendix
\section{The TABLE}\label{app:theTable}
We collect in Table \ref{tab:theTable} the numerical results from our suite of simulations. 
\startlongtable
\begin{deluxetable*}{ccccccccccccccccc}\label{tab:theTable}
\setlength{\tabcolsep}{3pt}
\tablehead{
\colhead{$\mathcal{M}$} & 
\colhead{$e$} & 
\colhead{viscosity} & 
\colhead{$\ell_0$} & 
\colhead{$\langle\dot{m}\rangle$} & 
\colhead{$\langle\dot{E}\rangle$} & 
\colhead{$\langle\dot{E}_g\rangle$} & 
\colhead{$\langle\dot{E}_a\rangle$} & 
\colhead{$\langle\dot{J}\rangle$} & 
\colhead{$\langle\dot{J}_g\rangle$} & 
\colhead{$\langle\dot{J}_a\rangle$} & 
\colhead{$\langle\dot{a}\rangle$} & 
\colhead{$\langle\dot{a}_g\rangle$} & 
\colhead{$\langle\dot{a}_a\rangle$} & 
\colhead{$\langle\dot{e}\rangle$} & 
\colhead{$\langle\dot{e}_g\rangle$} & 
\colhead{$\langle\dot{e}_a\rangle$} 
}
\startdata
10 & 0.003125 & $\nu\!=\!10^{-3}$ & 0.00 & 1.25 & 0.10 & 0.48 & -0.37 & 0.73 & 0.48 & 0.25 & 2.82 & 3.81 & -0.99 & -0.02 & -0.03 & 0.01 \\
10 & 0.0125 & $\nu\!=\!10^{-3}$ & 0.00 & 1.25 & 0.11 & 0.48 & -0.37 & 0.73 & 0.48 & 0.25 & 2.86 & 3.85 & -0.99 & -0.24 & -0.24 & 0.00 \\
10 & 0.025 & $\nu\!=\!10^{-3}$ & 0.00 & 0.99 & 0.11 & 0.49 & -0.37 & 0.74 & 0.49 & 0.25 & 2.89 & 3.88 & -0.99 & -0.48 & -0.49 & 0.00 \\
10 & 0.05 & $\nu\!=\!10^{-3}$ & 0.00 & 0.90 & 0.10 & 0.47 & -0.37 & 0.74 & 0.48 & 0.25 & 2.80 & 3.79 & -0.99 & -0.90 & -0.91 & 0.01 \\
10 & 0.1 & $\nu\!=\!10^{-3}$ & 0.00 & 1.00 & -0.31 & 0.06 & -0.37 & 0.13 & -0.12 & 0.25 & -0.49 & 0.49 & -0.99 & 7.40 & 7.37 & 0.03 \\
10 & 0.2 & $\nu\!=\!10^{-3}$ & 0.00 & 0.89 & -0.42 & -0.05 & -0.37 & -0.27 & -0.52 & 0.25 & -1.39 & -0.39 & -1.00 & 9.26 & 9.27 & -0.01 \\
10 & 0.3 & $\nu\!=\!10^{-3}$ & 0.00 & 0.68 & -1.01 & -0.63 & -0.38 & -1.33 & -1.57 & 0.24 & -6.07 & -5.02 & -1.05 & 12.31 & 12.39 & -0.08 \\
10 & 0.4 & $\nu\!=\!10^{-3}$ & 0.00 & 0.87 & -0.58 & -0.19 & -0.39 & -0.36 & -0.59 & 0.23 & -2.64 & -1.53 & -1.12 & 3.67 & 3.80 & -0.13 \\
10 & 0.45 & $\nu\!=\!10^{-3}$ & 0.00 & 1.21 & -0.16 & 0.23 & -0.40 & 0.63 & 0.41 & 0.22 & 0.69 & 1.85 & -1.16 & -1.75 & -1.60 & -0.15 \\
10 & 0.5 & $\nu\!=\!10^{-3}$ & 0.00 & 1.28 & -0.19 & 0.21 & -0.40 & 0.74 & 0.52 & 0.22 & 0.48 & 1.67 & -1.19 & -2.52 & -2.38 & -0.14 \\
10 & 0.6 & $\nu\!=\!10^{-3}$ & 0.00 & 1.26 & -0.35 & 0.06 & -0.41 & 0.74 & 0.54 & 0.20 & -0.80 & 0.45 & -1.25 & -2.78 & -2.65 & -0.14 \\
\hline
10 & 0.025 & $\nu\!=\!10^{-3}$ & 0.70 & 0.98 & 0.11 & 0.48 & -0.37 & 0.74 & 0.49 & 0.25 & 2.88 & 3.88 & -0.99 & -0.42 & -0.43 & 0.00 \\
10 & 0.05 & $\nu\!=\!10^{-3}$ & 0.70 & 0.98 & 0.10 & 0.48 & -0.37 & 0.74 & 0.49 & 0.25 & 2.83 & 3.82 & -0.99 & -0.78 & -0.79 & 0.01 \\
\hline
10 & 0.4 & $\nu\!=\!2.5\!\!\times\!\!10^{-4}$ & 0.00 & 1.32 & 0.09 & 0.47 & -0.38 & 0.65 & 0.42 & 0.23 & 2.71 & 3.72 & -1.02 & 0.07 & 0.10 & -0.03 \\
10 & 0.5 & $\nu\!=\!2.5\!\!\times\!\!10^{-4}$ & 0.00 & 1.42 & 0.10 & 0.48 & -0.38 & 0.76 & 0.55 & 0.22 & 2.82 & 3.87 & -1.05 & -0.92 & -0.87 & -0.05 \\
10 & 0.6 & $\nu\!=\!2.5\!\!\times\!\!10^{-4}$ & 0.00 & 1.40 & -0.13 & 0.26 & -0.39 & 0.74 & 0.54 & 0.20 & 0.94 & 2.04 & -1.11 & -1.83 & -1.77 & -0.06 \\
\hline
10 & 0.003125 & $\alpha\!=\!0.1$ & 0.00 & 1.17 & 0.16 & 0.54 & -0.37 & 0.79 & 0.54 & 0.25 & 3.30 & 4.29 & -0.99 & 0.11 & 0.08 & 0.03 \\
10 & 0.025 & $\alpha\!=\!0.1$ & 0.00 & 1.17 & 0.16 & 0.54 & -0.37 & 0.78 & 0.53 & 0.25 & 3.30 & 4.29 & -0.99 & 0.64 & 0.64 & -0.00 \\
10 & 0.05 & $\alpha\!=\!0.1$ & 0.00 & 1.17 & 0.15 & 0.53 & -0.37 & 0.77 & 0.51 & 0.25 & 3.23 & 4.21 & -0.99 & 0.89 & 0.90 & -0.00 \\
10 & 0.4 & $\alpha\!=\!0.1$ & 0.00 & 0.93 & -0.33 & 0.05 & -0.38 & -0.07 & -0.30 & 0.23 & -0.64 & 0.41 & -1.05 & 3.14 & 3.20 & -0.07 \\
10 & 0.45 & $\alpha\!=\!0.1$ & 0.00 & 1.26 & 0.02 & 0.41 & -0.39 & 0.69 & 0.46 & 0.22 & 2.19 & 3.27 & -1.08 & -0.86 & -0.78 & -0.08 \\
10 & 0.5 & $\alpha\!=\!0.1$ & 0.00 & 1.18 & 0.09 & 0.47 & -0.39 & 0.82 & 0.61 & 0.22 & 2.70 & 3.79 & -1.09 & -1.43 & -1.35 & -0.08 \\
\hline
20 & 0.0125 & $\nu\!=\!10^{-3}$ & 0.00 & 1.03 & -0.47 & -0.10 & -0.37 & 0.16 & -0.10 & 0.25 & -1.78 & -0.79 & -0.99 & -0.94 & -0.96 & 0.02 \\
20 & 0.025 & $\nu\!=\!10^{-3}$ & 0.00 & 1.02 & -0.46 & -0.09 & -0.37 & 0.16 & -0.09 & 0.25 & -1.69 & -0.70 & -0.99 & 0.72 & 0.70 & 0.02 \\
20 & 0.05 & $\nu\!=\!10^{-3}$ & 0.00 & 1.11 & -0.29 & 0.09 & -0.37 & 0.33 & 0.08 & 0.25 & -0.30 & 0.69 & -0.99 & 0.45 & 0.43 & 0.02 \\
20 & 0.1 & $\nu\!=\!10^{-3}$ & 0.00 & 1.13 & -0.21 & 0.16 & -0.37 & 0.38 & 0.13 & 0.25 & 0.30 & 1.29 & -0.98 & 1.41 & 1.37 & 0.04 \\
20 & 0.15 & $\nu\!=\!10^{-3}$ & 0.00 & 0.88 & -0.54 & -0.17 & -0.37 & -0.37 & -0.62 & 0.25 & -2.32 & -1.34 & -0.98 & 11.96 & 11.92 & 0.04 \\
20 & 0.2 & $\nu\!=\!10^{-3}$ & 0.00 & 0.64 & -1.29 & -0.91 & -0.37 & -1.84 & -2.08 & 0.25 & -8.30 & -7.31 & -0.99 & 23.32 & 23.30 & 0.02 \\
20 & 0.25 & $\nu\!=\!10^{-3}$ & 0.00 & 0.58 & -1.53 & -1.15 & -0.37 & -2.18 & -2.43 & 0.24 & -10.22 & -9.23 & -0.99 & 20.28 & 20.27 & 0.01 \\
20 & 0.3 & $\nu\!=\!10^{-3}$ & 0.00 & 0.69 & -1.05 & -0.67 & -0.37 & -1.36 & -1.60 & 0.24 & -6.38 & -5.39 & -0.99 & 12.19 & 12.18 & 0.01 \\
20 & 0.35 & $\nu\!=\!10^{-3}$ & 0.00 & 0.71 & -0.93 & -0.56 & -0.37 & -1.21 & -1.45 & 0.23 & -5.48 & -4.50 & -0.98 & 9.87 & 9.85 & 0.01 \\
20 & 0.4 & $\nu\!=\!10^{-3}$ & 0.00 & 1.10 & -0.12 & 0.26 & -0.38 & 0.19 & -0.04 & 0.23 & 1.05 & 2.08 & -1.03 & 2.49 & 2.53 & -0.04 \\
20 & 0.45 & $\nu\!=\!10^{-3}$ & 0.00 & 1.08 & -0.24 & 0.15 & -0.39 & 0.23 & 0.00 & 0.22 & 0.11 & 1.21 & -1.10 & 0.96 & 1.05 & -0.09 \\
20 & 0.5 & $\nu\!=\!10^{-3}$ & 0.00 & 1.24 & -0.28 & 0.13 & -0.41 & 0.50 & 0.28 & 0.22 & -0.20 & 1.05 & -1.25 & -1.38 & -1.18 & -0.19 \\
20 & 0.6 & $\nu\!=\!10^{-3}$ & 0.00 & 1.22 & -0.38 & 0.05 & -0.42 & 0.56 & 0.36 & 0.20 & -1.02 & 0.36 & -1.38 & -1.91 & -1.71 & -0.21 \\
\hline
25 & 0.4 & $\nu\!=\!10^{-3}$ & 0.00 & 0.78 & -0.68 & -0.30 & -0.38 & -0.95 & -1.18 & 0.23 & -3.41 & -2.38 & -1.03 & 8.28 & 8.31 & -0.04 \\
25 & 0.5 & $\nu\!=\!10^{-3}$ & 0.00 & 0.87 & -0.75 & -0.35 & -0.40 & -0.58 & -0.79 & 0.22 & -3.98 & -2.79 & -1.18 & 3.25 & 3.39 & -0.14 \\
25 & 0.6 & $\nu\!=\!10^{-3}$ & 0.00 & 0.90 & -0.85 & -0.42 & -0.43 & -0.35 & -0.55 & 0.20 & -4.83 & -3.40 & -1.43 & 0.91 & 1.13 & -0.23 \\
\hline
30 & 0.003125 & $\nu\!=\!10^{-3}$ & 0.00 & 0.82 & -1.11 & -0.73 & -0.37 & -0.48 & -0.73 & 0.25 & -6.87 & -5.87 & -0.99 & -0.07 & -0.08 & 0.01 \\
30 & 0.00625 & $\nu\!=\!10^{-3}$ & 0.00 & 0.82 & -1.13 & -0.76 & -0.37 & -0.51 & -0.76 & 0.25 & -7.07 & -6.07 & -0.99 & 0.24 & 0.23 & 0.01 \\
30 & 0.0125 & $\nu\!=\!10^{-3}$ & 0.00 & 0.82 & -1.16 & -0.78 & -0.37 & -0.54 & -0.79 & 0.25 & -7.27 & -6.27 & -0.99 & 1.81 & 1.80 & 0.01 \\
30 & 0.025 & $\nu\!=\!10^{-3}$ & 0.00 & 0.80 & -1.21 & -0.84 & -0.37 & -0.60 & -0.85 & 0.25 & -7.70 & -6.70 & -0.99 & 2.41 & 2.39 & 0.01 \\
30 & 0.05 & $\nu\!=\!10^{-3}$ & 0.00 & 0.76 & -1.43 & -1.06 & -0.37 & -0.84 & -1.09 & 0.25 & -9.44 & -8.45 & -0.99 & 2.51 & 2.48 & 0.03 \\
30 & 0.075 & $\nu\!=\!10^{-3}$ & 0.00 & 0.76 & -1.42 & -1.05 & -0.37 & -0.84 & -1.09 & 0.25 & -9.36 & -8.37 & -0.99 & 2.30 & 2.26 & 0.04 \\
30 & 0.1 & $\nu\!=\!10^{-3}$ & 0.00 & 0.55 & -2.00 & -1.63 & -0.37 & -2.38 & -2.63 & 0.25 & -14.04 & -13.05 & -0.99 & 39.94 & 39.91 & 0.03 \\
30 & 0.15 & $\nu\!=\!10^{-3}$ & 0.00 & 0.50 & -2.12 & -1.75 & -0.37 & -3.01 & -3.26 & 0.25 & -14.98 & -13.99 & -0.99 & 40.33 & 40.31 & 0.02 \\
30 & 0.2 & $\nu\!=\!10^{-3}$ & 0.00 & 0.51 & -1.94 & -1.57 & -0.37 & -2.79 & -3.03 & 0.25 & -13.52 & -12.53 & -0.99 & 29.38 & 29.38 & 0.00 \\
30 & 0.3 & $\nu\!=\!10^{-3}$ & 0.00 & 0.47 & -2.00 & -1.63 & -0.38 & -3.28 & -3.52 & 0.24 & -14.02 & -13.01 & -1.01 & 25.02 & 25.05 & -0.03 \\
30 & 0.4 & $\nu\!=\!10^{-3}$ & 0.00 & 0.47 & -2.08 & -1.69 & -0.39 & -3.39 & -3.62 & 0.23 & -14.62 & -13.50 & -1.12 & 18.91 & 19.03 & -0.12 \\
30 & 0.5 & $\nu\!=\!10^{-3}$ & 0.00 & 0.53 & -1.83 & -1.41 & -0.42 & -2.66 & -2.88 & 0.22 & -12.66 & -11.30 & -1.37 & 11.18 & 11.45 & -0.27 \\
30 & 0.6 & $\nu\!=\!10^{-3}$ & 0.00 & 0.66 & -1.33 & -0.87 & -0.46 & -1.48 & -1.68 & 0.20 & -8.65 & -6.96 & -1.69 & 4.87 & 5.23 & -0.36 \\
\hline
30 & 0.0125 & $\nu\!=\!10^{-3}$ & -0.54 & 0.94 & -1.18 & -0.80 & -0.37 & -0.56 & -0.81 & 0.25 & -7.42 & -6.43 & -0.99 & 1.60 & 1.59 & 0.01 \\
30 & 0.025 & $\nu\!=\!10^{-3}$ & -0.60 & 0.96 & -1.19 & -0.81 & -0.37 & -0.58 & -0.83 & 0.25 & -7.50 & -6.51 & -0.99 & 2.50 & 2.49 & 0.02 \\
30 & 0.5 & $\nu\!=\!10^{-3}$ & -1.70 & 0.80 & -1.75 & -1.32 & -0.43 & -2.53 & -2.75 & 0.22 & -11.99 & -10.53 & -1.46 & 10.78 & 11.12 & -0.34 \\
30 & 0.6 & $\nu\!=\!10^{-3}$ & -1.10 & 0.95 & -1.21 & -0.74 & -0.47 & -1.29 & -1.49 & 0.20 & -7.65 & -5.91 & -1.74 & 4.40 & 4.78 & -0.39 \\
\hline
35 & 0.4 & $\nu\!=\!10^{-3}$ & 0.00 & 0.34 & -3.31 & -2.92 & -0.39 & -5.80 & -6.03 & 0.23 & -24.52 & -23.38 & -1.14 & 30.56 & 30.70 & -0.14 \\
35 & 0.5 & $\nu\!=\!10^{-3}$ & 0.00 & 0.41 & -2.55 & -2.11 & -0.43 & -4.16 & -4.38 & 0.22 & -18.37 & -16.90 & -1.47 & 17.33 & 17.67 & -0.35 \\
35 & 0.6 & $\nu\!=\!10^{-3}$ & 0.00 & 0.55 & -1.74 & -1.27 & -0.48 & -2.45 & -2.65 & 0.20 & -11.95 & -10.15 & -1.80 & 8.30 & 8.72 & -0.42 \\
\hline
\enddata
\tablecomments{With the exception of the accretion rate itself, each reported time derivative as been normalized by $\langle\dot{m}\rangle/m$.}
\end{deluxetable*}

\end{document}